\documentclass[]{aa} 
\usepackage{epsfig, graphicx,longtable}
\usepackage{natbib}
\usepackage{times}

\bibpunct{(}{)}{;}{a}{}{,}

\begin{document}

%\thesaurus{06(08.01.1; 08.03.02; 08.16.2)}

%\titlerunning{} 
\title{Chemical abundances of planet-host stars\thanks{Based on 
               observations collected at the La Silla Observatory, 
               ESO (Chile), with the CORALIE spectrograph 
               at the 1.2-m Euler Swiss telescope and the FEROS spectrograph
               at the 1.52-m ESO telescope, with the VLT/UT2 
               Kueyen telescope (Paranal Observatory, ESO, Chile) using the 
               UVES spectrograph (Observing run 67.C-0206, in service 
               mode), with the 
               TNG and William Herschel Telescopes, both operated at the 
               island of La Palma, and with the ELODIE spectrograph at the
               1.93-m telescope at the Observatoire de Haute Provence.}}
\subtitle{Results for alpha and Fe-group elements}

\author{
      A. Bodaghee \inst{1}
 \and N.~C.~Santos \inst{1,2}
 \and G.~Israelian \inst{3}
 \and M.~Mayor \inst{1}
 }

\offprints{Nuno C. Santos, \email{Nuno.Santos@oal.ul.pt}}

\institute{
	Observatoire de Gen\`eve, 51 ch.  des 
	Maillettes, CH--1290 Sauverny, Switzerland
     \and
	Centro de Astronomia e Astrof{\'\i}sica da Universidade de Lisboa,
	Observat\'orio Astron\'omico de Lisboa, Tapada da Ajuda, 1349-018 
	Lisboa, Portugal
     \and
	Instituto de Astrof{\'\i}sica de Canarias, E-38200 
        La Laguna, Tenerife, Spain}
\date{Received / Accepted } 

\titlerunning{} 

%--------------------------------------------------------------------------

\abstract{
In this paper, we present a study of the abundances of Si, Ca, Sc, Ti, V, Cr, Mn, Co, and Ni 
in a large set of stars known to harbor giant planets,
as well as in a comparison sample of stars not known to have any planetary-mass
companions. We have checked for possible chemical differences between planet hosts 
and field stars without known planets. Our results show that overall, and for a given value of [Fe/H], the abundance trends for the planet hosts are nearly indistinguishable 
from those of the field stars. In general, the trends show no discontinuities, 
and the abundance distributions of stars with giant planets are high [Fe/H] 
extensions to the curves traced by the field dwarfs without planets. 
The only elements that might present slight differences between the two groups 
of stars are V, Mn, and to a lesser extent Ti and Co. 
We also use the available data to describe galactic chemical evolution trends
for the elements studied. When comparing the results with former studies, a few differences emerge for the high [Fe/H] tail of the distribution, a region
that is sampled with unprecedented detail in our analysis.
}

\maketitle

\section{Introduction}

Stars with planetary companions have been shown to be, on average, considerably
metal-rich when compared with stars in the solar neighborhood
\citep[e.g.][]{Gon97,Fur97,Gon98,Sad99,San00,Gon01,San01,San02}. The most recent results 
suggest indeed that the efficiency of planetary formation seems
to be strongly dependent on the metal content of the cloud that gave origin to
the star and planetary system.

Until now, however, most chemical studies of the planet hosts used iron as the reference element. 
The few systematic studies in the literature concerning
other metals \citep[][]{Gon97,Sad99,Gon00,San00,Sad01,Gon01,Smi01,Sad02} revealed a few possible 
(but not clear) anomalies. The situation concerning the light elements
\citep[in particular Li and Be -- ][]{Gar98,Del00,Gon00,Rya00,Gon01,Isr01,San02b,Red02,Isr03} is not
very different; the debate is just beginning to heat up and many questions remain open. 

In almost every case\footnote{With probably the only exception being the studies of beryllium
by \citet[][]{San02b}.}, the authors have been constrained to compare the results 
for the star-with-planet samples with other studies in the literature concerning stars without known 
planetary companions. This might have introduced undesirable
systematic errors, given that the different studies have not used the same set of spectral lines and 
model atmospheres to derive the stellar parameters and abundances. How important are the 
systematic differences between the various studies? And how can these differences lead to mistakes?
A systematic comparison between two groups of stars (with and without planetary companions)
is thus needed.

To try to fill (at least in part) this gap, we present in this paper a detailed and
uniform study of the elements Si, Ca, Sc, Ti, V, Cr, Mn, Co, and Ni in
a large sample of planet-host stars, and in a ``comparison'' volume-limited sample of
stars not known to have any planetary-mass companions. In Sect.\ref{sec:data}, we present
our samples as well as the chemical analysis, and 
in Sect.\ref{sec:comparing} we compare the abundances. The results seem to indicate that no clearly significant differences 
exist between the two groups of stars. In Sect.\ref{sec:galactic}, we use the current data to explore 
the galactic chemical evolution trends. Given the high metal content of many stars in our sample,
we could access the high [Fe/H] tail of the distributions with unprecedented detail. 
We conclude in Sect.\ref{sec:conclusions}.

\section{Data, atmospheric parameters, and chemical analysis}
\label{sec:data}

\subsection{The data}

In a series of recent papers \citep[][]{San00,San01,San02}, we have been gathering
spectra for most of the planet-host stars known today. This data has 
been used to derive precise and uniform stellar parameters for the
target stars, as well as accurate iron abundances. The results published to this point 
have been used to show that stars with planets are substantially metal-rich when
compared with average field dwarfs. 

The current paper employs the same spectra and stellar parameters derived in these works.
This allowed access to elemental abundances for 77 stars with low-mass companions (planetary 
or brown-dwarf candidates).

Spectra for the comparison sample used in this work were obtained 
with the main goal of deriving the metallicities for a large sample of
stars in a limited volume around the Sun and not known to harbor
any planetary-mass companions \citep[][]{San01}. With the exception
of \object{HD\,39091} (that was, in the meanwhile, found to have a brown-dwarf
candidate companion -- \citep[][]{Jon02}), the comparison sample used here consists of 
the remaining 42 objects in Table\,1 of \citet[][]{San01}. 
This sample is indeed perfect and appropriate for the current work, as
its stellar parameters and iron abundances have been derived using the same methods 
as those used for all the planet hosts analyzed in this paper. 

{We should caution that this comparison sample is built from a
list of stars that are surveyed for planets, but for which none have yet been
found. Of course, this does not mean that these stars do not have any
planetary mass companions at all (they might have e.g. very low mass and/or
long period planets which are more difficult to detect with radial-velocity surveys). 
However, the odds that planets similar to the ones found to date are present among
these stars are not very high.}

In the rest of the
paper, all the stellar parameters and [Fe/H] values have been 
taken from the uniform studies of \citet[][]{San00,San01,San02} for both samples. 
For more details on the reduction and analysis we refer to these works. 

\subsection{Chemical Analysis}

Abundances for all the elements studied here were derived
in standard Local Thermodynamic Equilibrium (LTE) using a revised version of the 
code MOOG \citep{Sne73}, and a grid of \citet{Kur93} ATLAS9 atmospheres.
For each element, a set of (weak) lines was chosen from the literature,
with wavelengths between about 5000 and 6800\AA\ (the usual spectral domain of
our data). Then, using the Kurucz Solar Atlas \citep[][]{Kur84}, we verified each line
to check for possible blends. Only isolated lines were taken; for these,
the solar equivalent widths (EW) were measured.

\begin{table}[thbp] \centering
	\caption{ Spectral lines of the elements used in this experiment. 
Col.\,1: wavelength (in \AA ngstroms). Col.\,2: excitation energy of the lower energy level in the transition (in eV). Col.\,3: oscillator strengths based on an inverse solar analysis. }
   	\begin{tabular}{ccrcccr}	\hline
   	\noalign{\smallskip}
   	\noalign{\smallskip}
	$\lambda$ & $\chi_{l}$ & $\log{gf}$ & &  $\lambda$ & $\chi_{l}$ & $\log{gf}$\\
	\noalign{\smallskip}
	\noalign{\smallskip}
	\hline
	\noalign{\smallskip}
	\multicolumn{3}{l}{\textbf{Si} I ; $\log \epsilon_{\circ} = 7.55$  A $=$ 14} & & 6261.11 & 1.43 & $-$0.4881 \\
	5665.56 & 4.92 & $-$1.9788 & & 6303.76 & 1.44 & $-$1.6003 \\
	5690.43 & 4.93 & $-$1.7878 & & 6312.24 & 1.46 & $-$1.5850 \\
	5701.10 & 4.93 & $-$2.0227 & & \multicolumn{3}{l}{\textbf{V} I ; $\log \epsilon_{\circ} = 4.00$  A $=$ 23} \\
	5772.14 & 5.08 & $-$1.6179 & & 5727.05 & 1.08 & $-$0.0004 \\
	5793.09 & 4.93 & $-$1.9134 & & 6090.21 & 1.08 & $-$0.1549 \\
	5948.55 & 5.08 & $-$1.1135 & & 6216.35 & 0.28 & $-$0.8996 \\
	6125.02 & 5.61 & $-$1.5157 & & 6452.31 & 1.19 & $-$0.8239 \\
	6142.49 & 5.62 & $-$1.4788 & & 6531.42 & 1.22 & $-$0.9208 \\
	6145.02 & 5.61 & $-$1.4012 & & \multicolumn{3}{l}{\textbf{Cr} I ; $\log \epsilon_{\circ} = 5.67$  A $=$ 24} \\
	6155.15 & 5.62 & $-$0.7520 & & 5304.18 & 3.46 & $-$0.6777 \\
	6721.86 & 5.86 & $-$1.0872 & & 5312.86 & 3.45 & $-$0.5850 \\
	\multicolumn{3}{l}{\textbf{Ca} I ; $\log \epsilon_{\circ} = 6.36$ A $=$ 20} & & 5318.77 & 3.44 & $-$0.7099 \\
	5512.98 & 2.93 & $-$0.4377 & & 5480.51 & 3.50 & $-$0.8268 \\
	5581.97 & 2.52 & $-$0.6536 & & 5574.39 & 4.45 & $-$0.4814 \\
	5590.12 & 2.52 & $-$0.7077 & & 5783.07 & 3.32 & $-$0.4034 \\
	5867.56 & 2.93 & $-$1.5900 & & 5783.87 & 3.32 & $-$0.1487 \\
	6161.29 & 2.52 & $-$1.2182 & & 5787.92 & 3.32 & $-$0.1067 \\
	6166.44 & 2.52 & $-$1.1163 & & \multicolumn{3}{l}{\textbf{Mn} I ; $\log \epsilon_{\circ} = 5.39$ A $=$ 25} \\
	6169.05 & 2.52 & $-$0.7328 & & 5388.50 & 3.37 & $-$1.6289 \\
	6169.56 & 2.52 & $-$0.4436 & & 5399.47 & 3.85 & $-$0.0969 \\
	6449.82 & 2.52 & $-$0.6289 & & 6440.93 & 3.77 & $-$1.2518 \\
	6455.60 & 2.52 & $-$1.3736 & & \multicolumn{3}{l}{\textbf{Co} I ; $\log \epsilon_{\circ} = 4.92$ A $=$ 27} \\
	\multicolumn{3}{l}{\textbf{Sc} II ; $\log \epsilon_{\circ} = 3.10$ A $=$ 21} & & 5301.04 & 1.71 & $-$1.9318 \\
	5239.82 & 1.45 & $-$0.7594 & & 5312.65 & 4.21 & $-$0.0199 \\
	5318.36 & 1.36 & $-$1.6989 & & 5483.36 & 1.71 & $-$1.2182 \\
	5526.82 & 1.77 & 0.14612 & & 6455.00 & 3.63 & $-$0.2839 \\
	6245.62 & 1.51 & $-$1.0409 & & 6632.44 & 2.28 & $-$1.8827 \\
	6300.69 & 1.51 & $-$1.9586 & & \multicolumn{3}{l}{\textbf{Ni} I ; $\log \epsilon_{\circ} = 6.25$ A $=$ 28} \\
	6320.84 & 1.50 & $-$1.8386 & & 5578.72 & 1.68 & $-$2.6517 \\
	6604.60 & 1.36 & $-$1.1611 & & 5587.86 & 1.93 & $-$2.3819 \\
	\multicolumn{3}{l}{\textbf{Ti} I ; $\log \epsilon_{\circ} = 4.99$ A $=$ 22} & & 5682.20 & 4.10 & $-$0.3872 \\
	5471.20 & 1.44 & $-$1.5528 & & 5694.99 & 4.09 & $-$0.6020 \\
	5474.23 & 1.46 & $-$1.3615 & & 5805.22 & 4.17 & $-$0.5767 \\
	5490.15 & 1.46 & $-$0.9829 & & 5847.00 & 1.68 & $-$3.4089 \\
	5866.46 & 1.07 & $-$0.8386 & & 6086.28 & 4.26 & $-$0.4436 \\
	6091.18 & 2.27 & $-$0.4559 & & 6111.07 & 4.09 & $-$0.8013 \\
	6126.22 & 1.07 & $-$0.4145 & & 6128.98 & 1.68 & $-$3.3665 \\
	6258.11 & 1.44 & $-$0.4365 & & 6130.14 & 4.26 & $-$0.9469 \\
	\noalign{\smallskip}
	\hline
   	\end{tabular}
\label{tab1}
\end{table}

\begin{table*}[thbp] \centering
        \caption{ Sensitivity of the derived abundances to changes of 0.10 dex in
metallicity, 0.15 dex in gravity, 0.10 km\,s$^{-1}$ in microturbulence, and
50\,K in effective temperature for the K dwarf \object{HD\,50281\,A}, for a 
solar-type star (\object{HD\,43162}), and for the late-F dwarf \object{HD\,10647}. }
        \label{Tab4}
        \begin{tabular}{lrrrrrrrrr}     \hline
        \noalign{\smallskip}
        \noalign{\smallskip}
        Star & Si & Ca & Sc & Ti & V & Cr & Mn & Co & Ni \\
        \noalign{\smallskip}
        \noalign{\smallskip}
        \hline
        \noalign{\smallskip}
        \noalign{\smallskip}
        \textbf{HD 50281A} & \multicolumn{9}{c}{ ($[$Fe$/$H$]$\,;\,$\log\,g$\,;\,$\xi_{t}$\,;\,T$_{\mathrm{eff}}$) = (0.07\,;\,4.75\,;\,0.85\,;\,4790)} \\
        \noalign{\smallskip}
        $\Delta [$Fe$/$H$] = +$0.10 dex & 0.03 & 0.02 & 0.04 & 0.01 & 0.02 & 0.02 & 0.02 & 0.03 & 0.03  \\
        $\Delta \log\,g  = +$0.15 dex & 0.03 & 0.02 & 0.06 & $-$0.02 & $-$0.03 & $-$0.01 & 0.00 & 0.03 & 0.02  \\
        $\Delta \xi_{t}  = +$0.10 km$\cdot$s$^{-1}$ & 0.00 & $-$0.01 & $-$0.01 & $-$0.03 & $-$0.04 & $-$0.01 & $-$0.01 & $-$0.02 & $-$0.02  \\
        $\Delta$ T$_\mathrm{eff}  = +$50 K & $-$0.02 & 0.05 & $-$0.01 & 0.06 & 0.06 & 0.03 & 0.01 & 0.01 & $-$0.01  \\
        \noalign{\smallskip}
        \noalign{\smallskip}
        \noalign{\smallskip}
        \textbf{HD 43162} & \multicolumn{9}{c}{ ($[$Fe$/$H$]$\,;\,$\log\,g$\,;\,$\xi_{t}$\,;\,T$_\mathrm{eff}$) = ($-$0.02\,;\,4.57\,;\,1.36\,;\,5630)} \\
        \noalign{\smallskip}
        $\Delta [$Fe$/$H$] = +$0.10 dex & 0.01 & 0.00 & 0.03 & 0.00 & $-$0.01 & 0.00 & 0.00 & 0.01 & 0.01  \\
        $\Delta \log\,g  = +$0.15 dex & 0.01 & $-$0.04 & 0.06 & 0.00 & $-$0.01 & $-$0.01 & 0.00 & 0.01 & 0.01  \\
        $\Delta \xi_{t}  = +$0.10 km$\cdot$s$^{-1}$ & $-$0.01 & $-$0.03 & $-$0.01 & $-$0.01 & $-$0.02 & $-$0.01 & $-$0.01 & $-$0.01 & $-$0.01  \\
        $\Delta$ T$_\mathrm{eff}  = +$50 K & 0.00 & 0.03 & 0.00 & 0.05 & 0.05 & 0.03 & 0.03 & 0.04 & 0.03 \\
        \noalign{\smallskip}
        \noalign{\smallskip}
        \noalign{\smallskip}
        \textbf{HD 10647} & \multicolumn{9}{c}{ ($[$Fe$/$H$]$\,;\,$\log\,g$\,;\,$\xi_{t}$\,;\,T$_\mathrm{eff}$) = ($-$0.03\,;\,4.45\,;\,1.31\,;\,6130) } \\
        \noalign{\smallskip}
        $\Delta [$Fe$/$H$] = +$0.10 dex & 0.00 & $-$0.01 & 0.02 & 0.00 & 0.00 & 0.00 & $-$0.01 & 0.00 & 0.00 \\
        $\Delta \log\,g  = +$0.15 dex & 0.00 & $-$0.02 & 0.06 & 0.00 & 0.00 & 0.00 & $-$0.01 & 0.00 & 0.00  \\
        $\Delta \xi_{t}  = +$0.10 km$\cdot$s$^{-1}$ & $-$0.01 & $-$0.02 & $-$0.02 & 0.00 & 0.00 & $-$0.01 & $-$0.01 & $-$0.01 & $-$0.01  \\
        $\Delta$ T$_\mathrm{eff}  = +$50 K & 0.01 & 0.03 & 0.01 & 0.04 & 0.05 & 0.03 & 0.02 & 0.04 & 0.03 \\
        \noalign{\smallskip}
        \hline
        \end{tabular}
\end{table*}

Using these measured EW values and a solar atmosphere with 
(T$_{\mathrm{eff}}$,$\log{g}$,$\xi_t$)=(5770\,K,4.44\,dex,1.00\,km\,s$^{-1}$), we have derived
$\log{gf}$ values for the individual lines from an inverted solar analysis (using
the {\tt ewfind} driver in MOOG). The solar abundances were taken from \citet[][]{And89}. In order to 
do the subsequent chemical analysis for our targets in a reasonable 
amount of time, we reduced the number of lines per element
to a maximum of $\sim$10. The final list of lines used is presented in Table\,\ref{tab1}.

For each line in our targets, the EW was measured using the IRAF\footnote{IRAF 
is distributed by National Optical Astronomy Observatories, operated 
by the Association of Universities for Research in Astronomy, Inc., 
under contract with the National Science Foundation, U.S.A.} {\tt splot} tool, and
the abundances were computed using MOOG with the {\tt abfind} driver. 
The average of the abundances for the lines of a given element was
then considered. In Tables\,\ref{tab2}, \ref{tab2b}, \ref{tab3}, and \ref{tab3b} we summarize the
derived abundances for all stars with and without planetary-mass
companions. The uncertainties in the tables denote the rms around the
mean. The number of lines used in each case is also noted.

We remark that in particular cases, some lines were eliminated from the
analysis when the quality of the spectrum in the region of interest was not
good enough to permit a reliable EW measurement (e.g. lower than usual S/N or
the presence of cosmic-rays).

\subsection{Uncertainties}

Uncertainties can sway the abundances in various ways. For example, 
errors can affect individual lines with random errors in the EWs, oscillator strengths 
and damping constants. Systematic errors in the EWs can arise from unnoticed blends or a 
poor location of the continuum. These errors are hard to spot, but they are
minimized thanks to the normally high quality of our data. 
Atmospheric parameter uncertainties should be the primary source of 
abundance error in a species with many lines, whereas inaccuracies in the EWs are eventually more 
important when only a few lines are available. 

Assuming perturbations of 0.10\,dex in the overall metallicity 
(scaled with [Fe/H])\footnote{The usual errors in this quantity for the stars 
in our sample are smaller than this value \citep[][]{San01}.}, 0.15 dex 
in $\log g$, 0.10 km\,s$^{-1}$ in $\xi_{t}$, and 50~K in effective temperatures (usual 
values for our sample), this leads to a total typical uncertainty of about 0.05 dex in the [X/H] 
abundance ratios (see Table~\ref{Tab4}) using the lines listed in Table\,\ref{tab1}.
Adding quadratically to the abundance dispersions for each element, we estimate that 
the errors in the abundances derived here are usually lower than 0.10\,dex\footnote{In a relative and
not absolute sense.}.

\begin{table*}[t]
	\caption{Atmospheric parameters \citep[taken from][]{San01,San02} and abundances derived for 
	Si, Ca, Sc, and Ti for the stars with giant planets studied in this paper. All the abundances
	are expressed as [X/H]=$\log{N(\mathrm{X})/N(\mathrm{H})}$+12. The number of spectral lines used is given by $n$, while
	$\sigma$ denotes the rms around the average.}
	\label{tab2}
	\begin{scriptsize}
   	\begin{tabular}{lcccrrcrrcrrcrrcr}	\hline
   	\noalign{\smallskip}
	Star & T$_\mathrm{eff}$ & $\log g$ & $\xi_{t}$ & Fe & Si & $\sigma$ & $n$ & Ca & $\sigma$ & $n$ & Sc & $\sigma$ & $n$ & Ti & $\sigma$ & $n$ \\
	\noalign{\smallskip}
	\hline
	\noalign{\smallskip}
	HD 142 & 6290 & 4.38 & 1.91 & 0.11 & 0.14 & 0.04 & 10 & 0.08 & 0.08 & 10 & 0.10 & 0.07 & 4 & 0.03 & 0.04 & 2 \\
	HD 1237 & 5555 & 4.65 & 1.50 & 0.11 & 0.06 & 0.06 & 11 & 0.09 & 0.07 & 10 & 0.08 & 0.09 & 5 & 0.12 & 0.06 & 9 \\
	HD 2039 & 5990 & 4.56 & 1.24 & 0.34 & 0.33 & 0.03 & 11 & 0.29 & 0.05 & 10 & 0.36 & 0.05 & 5 & 0.33 & 0.04 & 8 \\
	HD 4203 & 5650 & 4.38 & 1.15 & 0.40 & 0.44 & 0.05 & 11 & 0.33 & 0.07 & 10 & 0.48 & 0.07 & 6 & 0.39 & 0.09 & 10 \\
	HD 4208 & 5625 & 4.54 & 0.95 & $-$0.23 & $-$0.20 & 0.05 & 11 & $-$0.20 & 0.03 & 10 & $-$0.18 & 0.08 & 5 & $-$0.17 & 0.07 & 10 \\
	HD 6434 & 5790 & 4.56 & 1.40 & $-$0.55 & $-$0.34 & 0.05 & 11 & $-$0.40 & 0.04 & 10 & $-$0.33 & 0.08 & 5 & $-$0.32 & 0.06 & 6 \\
	HD 8574 & 6080 & 4.41 & 1.25 & 0.05 & $-$0.02 & 0.04 & 11 & 0.01 & 0.08 & 6 & 0.12 & 0.07 & 4 & $-$0.05 & 0.04 & 3 \\
	HD 9826 & 6120 & 4.07 & 1.50 & 0.10 & 0.12 & 0.04 & 9 & 0.16 & 0.08 & 5 & 0.08 & 0.09 & 3 & 0.08 & 0.00 & 1 \\
	HD 10697 & 5665 & 4.18 & 1.19 & 0.14 & 0.14 & 0.02 & 11 & 0.09 & 0.04 & 7 & 0.23 & 0.09 & 6 & 0.12 & 0.03 & 9 \\
	HD 12661 & 5715 & 4.49 & 1.09 & 0.36 & 0.35 & 0.05 & 11 & 0.23 & 0.12 & 9 & 0.49 & 0.13 & 6 & 0.30 & 0.04 & 7 \\
	HD 13445 & 5205 & 4.70 & 0.82 & $-$0.20 & $-$0.14 & 0.05 & 11 & $-$0.19 & 0.07 & 9 & $-$0.12 & 0.12 & 6 & $-$0.01 & 0.08 & 10 \\
	HD 16141 & 5805 & 4.28 & 1.37 & 0.15 & 0.10 & 0.03 & 11 & 0.09 & 0.03 & 10 & 0.22 & 0.09 & 7 & 0.13 & 0.04 & 9 \\
	HD 17051 & 6225 & 4.65 & 1.20 & 0.25 & 0.20 & 0.04 & 11 & 0.23 & 0.05 & 10 & 0.24 & 0.06 & 6 & 0.20 & 0.08 & 6 \\
	HD 19994 & 6175 & 4.14 & 1.52 & 0.21 & 0.22 & 0.08 & 11 & 0.22 & 0.07 & 8 & 0.20 & 0.12 & 4 & 0.15 & 0.05 & 4 \\
	HD 20367 & 6100 & 4.55 & 1.31 & 0.14 & 0.06 & 0.08 & 11 & 0.05 & 0.06 & 10 & 0.01 & 0.11 & 5 & $-$0.01 & 0.14 & 6 \\
	HD 22049 & 5135 & 4.70 & 1.14 & $-$0.07 & $-$0.10 & 0.05 & 11 & $-$0.10 & 0.07 & 9 & $-$0.07 & 0.16 & 5 & 0.00 & 0.04 & 10 \\
	HD 23079 & 5945 & 4.44 & 1.21 & $-$0.11 & $-$0.14 & 0.05 & 11 & $-$0.08 & 0.05 & 10 & $-$0.12 & 0.13 & 7 & $-$0.11 & 0.05 & 7 \\
	HD 23596 & 6125 & 4.29 & 1.32 & 0.32 & 0.30 & 0.03 & 11 & 0.27 & 0.05 & 9 & 0.33 & 0.01 & 5 & 0.28 & 0.05 & 9 \\
	HD 27442 & 4890 & 3.89 & 1.24 & 0.42 & 0.46 & 0.13 & 11 & 0.18 & 0.09 & 8 & 0.57 & 0.12 & 6 & 0.39 & 0.09 & 9 \\
	HD 28185 & 5705 & 4.59 & 1.09 & 0.24 & 0.23 & 0.04 & 11 & 0.16 & 0.04 & 10 & 0.35 & 0.09 & 6 & 0.23 & 0.05 & 8 \\
	HD 30177 & 5590 & 4.45 & 1.07 & 0.39 & 0.41 & 0.07 & 11 & 0.23 & 0.07 & 10 & 0.52 & 0.20 & 7 & 0.32 & 0.09 & 10 \\
	HD 33636 & 5990 & 4.68 & 1.22 & $-$0.05 & $-$0.06 & 0.04 & 11 & $-$0.04 & 0.03 & 10 & 0.04 & 0.09 & 5 & $-$0.03 & 0.08 & 7 \\
	HD 37124 & 5565 & 4.62 & 0.90 & $-$0.37 & $-$0.24 & 0.06 & 11 & $-$0.28 & 0.05 & 10 & $-$0.16 & 0.10 & 5 & $-$0.14 & 0.05 & 8 \\
	HD 38529 & 5675 & 4.01 & 1.39 & 0.39 & 0.39 & 0.05 & 11 & 0.32 & 0.05 & 10 & 0.43 & 0.04 & 6 & 0.37 & 0.05 & 10 \\
	HD 39091 & 5995 & 4.48 & 1.30 & 0.09 & 0.08 & 0.02 & 11 & 0.05 & 0.02 & 10 & 0.13 & 0.05 & 7 & 0.05 & 0.04 & 9 \\
	HD 46375 & 5315 & 4.54 & 1.11 & 0.21 & 0.21 & 0.06 & 11 & 0.14 & 0.10 & 10 & 0.21 & 0.08 & 5 & 0.28 & 0.07 & 10 \\
	HD 50554 & 6050 & 4.59 & 1.19 & 0.02 & $-$0.01 & 0.06 & 11 & 0.01 & 0.06 & 10 & 0.10 & 0.06 & 5 & 0.03 & 0.05 & 6 \\
	HD 52265 & 6098 & 4.29 & 1.31 & 0.24 & 0.22 & 0.04 & 11 & 0.21 & 0.05 & 10 & 0.21 & 0.06 & 5 & 0.18 & 0.03 & 6 \\
	HD 74156 & 6105 & 4.40 & 1.36 & 0.15 & 0.13 & 0.03 & 11 & 0.12 & 0.07 & 10 & 0.19 & 0.00 & 4 & 0.10 & 0.03 & 7 \\
	HD 75289 & 6135 & 4.43 & 1.50 & 0.27 & 0.22 & 0.04 & 11 & 0.24 & 0.04 & 10 & 0.27 & 0.10 & 5 & 0.22 & 0.03 & 6 \\
	HD 75732 & 5307 & 4.58 & 1.06 & 0.35 & 0.38 & 0.06 & 11 & 0.19 & 0.09 & 10 & 0.40 & 0.10 & 6 & 0.38 & 0.11 & 10 \\
	HD 80606 & 5570 & 4.56 & 1.11 & 0.34 & 0.36 & 0.04 & 11 & 0.22 & 0.09 & 10 & 0.40 & 0.11 & 6 & 0.36 & 0.07 & 10 \\
	HD 82943 & 6025 & 4.54 & 1.12 & 0.32 & 0.31 & 0.07 & 11 & 0.24 & 0.03 & 10 & 0.30 & 0.05 & 6 & 0.24 & 0.05 & 10 \\
	HD 83443 & 5500 & 4.50 & 1.12 & 0.39 & 0.44 & 0.07 & 11 & 0.25 & 0.10 & 10 & 0.41 & 0.09 & 6 & 0.38 & 0.07 & 10 \\
	HD 92788 & 5820 & 4.60 & 1.12 & 0.34 & 0.32 & 0.03 & 11 & 0.25 & 0.05 & 10 & 0.44 & 0.11 & 7 & 0.35 & 0.03 & 10 \\
	HD 95128 & 5925 & 4.45 & 1.24 & 0.05 & 0.01 & 0.08 & 11 & 0.01 & 0.03 & 6 & 0.09 & 0.04 & 6 & 0.04 & 0.03 & 10 \\
	HD 106252 & 5890 & 4.40 & 1.06 & $-$0.01 & $-$0.07 & 0.04 & 11 & $-$0.06 & 0.03 & 10 & $-$0.04 & 0.10 & 7 & $-$0.10 & 0.06 & 9 \\
	HD 108147 & 6265 & 4.59 & 1.40 & 0.20 & 0.14 & 0.03 & 11 & 0.14 & 0.02 & 10 & 0.20 & 0.08 & 4 & 0.10 & 0.03 & 4 \\
	HD 108874 & 5615 & 4.58 & 0.93 & 0.25 & 0.21 & 0.04 & 11 & 0.12 & 0.05 & 8 & 0.36 & 0.10 & 6 & 0.23 & 0.09 & 9 \\
	HD 114386 & 4875 & 4.69 & 0.63 & 0.00 & 0.00 & 0.06 & 11 & $-$0.06 & 0.13 & 10 & 0.11 & 0.07 & 4 & 0.20 & 0.09 & 10 \\
	HD 114729 & 5820 & 4.20 & 1.03 & $-$0.26 & $-$0.24 & 0.03 & 10 & $-$0.23 & 0.05 & 8 & $-$0.06 & 0.11 & 6 & $-$0.18 & 0.04 & 6 \\
	HD 114762 & 5870 & 4.25 & 1.28 & $-$0.72 & $-$0.57 & 0.02 & 8 & $-$0.57 & 0.05 & 10 & $-$0.61 & 0.06 & 7 & $-$0.58 & 0.08 & 9 \\
	HD 114783 & 5160 & 4.75 & 0.79 & 0.16 & 0.14 & 0.05 & 10 & $-$0.03 & 0.10 & 6 & 0.27 & 0.05 & 6 & 0.25 & 0.08 & 10 \\
	HD 117176 & 5530 & 4.05 & 1.08 & $-$0.05 & $-$0.07 & 0.09 & 11 & $-$0.06 & 0.03 & 7 & 0.02 & 0.09 & 7 & $-$0.03 & 0.03 & 10 \\
	HD 121504 & 6090 & 4.73 & 1.35 & 0.17 & 0.12 & 0.03 & 11 & 0.09 & 0.03 & 10 & 0.19 & 0.09 & 5 & 0.17 & 0.04 & 6 \\
	HD 128311 & 4950 & 4.80 & 1.00 & 0.10 & 0.07 & 0.07 & 10 & $-$0.04 & 0.12 & 8 & 0.15 & 0.13 & 5 & 0.16 & 0.09 & 9 \\
	HD 130322 & 5430 & 4.62 & 0.92 & 0.06 & 0.02 & 0.08 & 10 & 0.01 & 0.06 & 6 & 0.09 & 0.12 & 7 & 0.11 & 0.04 & 10 \\
	HD 134987 & 5780 & 4.45 & 1.06 & 0.32 & 0.32 & 0.03 & 10 & 0.23 & 0.03 & 6 & 0.45 & 0.11 & 7 & 0.31 & 0.03 & 10 \\
	HD 136118 & 6175 & 4.18 & 1.61 & $-$0.06 & $-$0.08 & 0.06 & 10 & $-$0.03 & 0.08 & 6 & $-$0.13 & 0.13 & 7 & $-$0.15 & 0.12 & 5 \\
	HD 137759 & 4750 & 3.15 & 1.78 & 0.09 & 0.12 & 0.06 & 9 & $-$0.09 & 0.14 & 6 & 0.28 & 0.05 & 5 & 0.18 & 0.13 & 9 \\
	HD 141937 & 5925 & 4.62 & 1.16 & 0.11 & 0.15 & 0.04 & 11 & 0.10 & 0.03 & 10 & 0.18 & 0.13 & 7 & 0.11 & 0.07 & 10 \\
	HD 143761 & 5835 & 4.40 & 1.29 & $-$0.21 & $-$0.15 & 0.04 & 10 & $-$0.17 & 0.03 & 6 & $-$0.05 & 0.06 & 7 & $-$0.11 & 0.06 & 10 \\
	HD 145675 & 5255 & 4.40 & 0.68 & 0.51 & 0.49 & 0.09 & 10 & 0.29 & 0.10 & 6 & 0.51 & 0.13 & 6 & 0.44 & 0.08 & 8 \\
	HD 147513 & 5880 & 4.58 & 1.17 & 0.07 & $-$0.01 & 0.03 & 11 & 0.08 & 0.04 & 10 & 0.04 & 0.10 & 6 & 0.03 & 0.03 & 9 \\
	HD 150706 & 6000 & 4.62 & 1.16 & 0.01 & $-$0.06 & 0.03 & 10 & $-$0.03 & 0.04 & 8 & $-$0.04 & 0.08 & 5 & $-$0.05 & 0.07 & 8 \\
	HD 160691 & 5820 & 4.44 & 1.23 & 0.33 & 0.32 & 0.03 & 11 & 0.22 & 0.03 & 10 & 0.38 & 0.04 & 6 & 0.31 & 0.04 & 10 \\
	HD 162020 & 4830 & 4.76 & 0.72 & 0.01 & $-$0.08 & 0.07 & 8 & $-$0.15 & 0.11 & 8 & $-$0.12 & 0.22 & 4 & 0.09 & 0.09 & 10 \\
	HD 168443 & 5600 & 4.30 & 1.18 & 0.06 & 0.08 & 0.05 & 10 & 0.02 & 0.04 & 6 & 0.19 & 0.03 & 4 & 0.13 & 0.02 & 9 \\
	HD 168746 & 5610 & 4.50 & 1.02 & $-$0.06 & 0.01 & 0.04 & 11 & $-$0.03 & 0.06 & 10 & 0.09 & 0.15 & 5 & 0.08 & 0.04 & 8 \\
	HD 169830 & 6300 & 4.04 & 1.37 & 0.22 & 0.15 & 0.03 & 11 & 0.18 & 0.08 & 10 & 0.15 & 0.12 & 6 & 0.09 & 0.02 & 6 \\
	HD 177830 & 4840 & 3.60 & 1.18 & 0.32 & 0.35 & 0.06 & 8 & 0.17 & 0.11 & 8 & 0.38 & 0.10 & 5 & 0.33 & 0.12 & 9 \\
	HD 179949 & 6235 & 4.41 & 1.38 & 0.21 & 0.16 & 0.05 & 8 & 0.20 & 0.07 & 8 & 0.11 & 0.07 & 4 & 0.10 & 0.02 & 3 \\
	HD 186427 & 5765 & 4.46 & 1.03 & 0.09 & 0.04 & 0.06 & 10 & 0.04 & 0.02 & 6 & 0.16 & 0.10 & 7 & 0.06 & 0.04 & 9 \\
	HD 187123 & 5855 & 4.48 & 1.10 & 0.14 & 0.06 & 0.07 & 11 & 0.08 & 0.02 & 6 & 0.15 & 0.11 & 7 & 0.08 & 0.05 & 10 \\
	HD 190228 & 5340 & 3.99 & 1.11 & $-$0.24 & $-$0.23 & 0.02 & 11 & $-$0.24 & 0.03 & 6 & $-$0.12 & 0.08 & 7 & $-$0.15 & 0.02 & 9 \\
	HD 190360 & 5590 & 4.48 & 1.06 & 0.25 & 0.26 & 0.04 & 11 & 0.14 & 0.08 & 8 & 0.41 & 0.13 & 6 & 0.28 & 0.05 & 8 \\
	HD 192263 & 4995 & 4.76 & 0.90 & 0.04 & 0.02 & 0.07 & 10 & $-$0.06 & 0.11 & 10 & 0.09 & 0.16 & 5 & 0.11 & 0.09 & 10 \\
	HD 195019 & 5840 & 4.36 & 1.24 & 0.08 & 0.04 & 0.02 & 11 & 0.05 & 0.03 & 10 & 0.15 & 0.08 & 6 & 0.10 & 0.04 & 8 \\
	HD 196050 & 5905 & 4.41 & 1.40 & 0.21 & 0.23 & 0.03 & 11 & 0.13 & 0.02 & 10 & 0.33 & 0.08 & 6 & 0.17 & 0.06 & 10 \\
	HD 202206 & 5765 & 4.75 & 0.99 & 0.37 & 0.30 & 0.03 & 11 & 0.24 & 0.03 & 10 & 0.45 & 0.04 & 4 & 0.28 & 0.04 & 7 \\
	HD 209458 & 6120 & 4.56 & 1.37 & 0.02 & 0.01 & 0.03 & 9 & 0.01 & 0.02 & 10 & 0.07 & 0.06 & 7 & $-$0.02 & 0.06 & 8 \\
	HD 210277 & 5570 & 4.45 & 1.08 & 0.22 & 0.23 & 0.04 & 11 & 0.20 & 0.08 & 10 & 0.23 & 0.10 & 6 & 0.27 & 0.04 & 10 \\
	HD 213240 & 5975 & 4.32 & 1.30 & 0.16 & 0.12 & 0.04 & 11 & 0.09 & 0.04 & 10 & 0.24 & 0.08 & 6 & 0.11 & 0.04 & 6 \\
	HD 216435 & 5905 & 4.16 & 1.26 & 0.22 & 0.19 & 0.04 & 11 & 0.14 & 0.06 & 10 & 0.20 & 0.05 & 5 & 0.11 & 0.04 & 7 \\
	HD 216437 & 5875 & 4.38 & 1.30 & 0.25 & 0.22 & 0.03 & 11 & 0.16 & 0.03 & 10 & 0.33 & 0.12 & 7 & 0.17 & 0.03 & 10 \\
	HD 217014 & 5805 & 4.51 & 1.22 & 0.21 & 0.21 & 0.04 & 11 & 0.12 & 0.03 & 9 & 0.29 & 0.09 & 7 & 0.18 & 0.05 & 9 \\
	HD 217107 & 5658 & 4.43 & 1.08 & 0.39 & 0.37 & 0.04 & 11 & 0.28 & 0.07 & 10 & 0.46 & 0.13 & 5 & 0.34 & 0.04 & 10 \\
	HD 222582 & 5850 & 4.58 & 1.06 & 0.06 & 0.00 & 0.05 & 9 & $-$0.10 & 0.15 & 9 & 0.08 & 0.10 & 5 & $-$0.03 & 0.10 & 4 \\
	\noalign{\smallskip}
	\hline
   	\end{tabular}
	\end{scriptsize}
\end{table*}

\begin{table*}[thbp]
	\caption{ Same as Table\,\ref{tab2} but for V, Cr, Mn, Co, and Ni.}
	\label{tab2b}
	\begin{scriptsize}
   	\begin{tabular}{lcrcrrcrrcrrcrrcr}	\hline
   	\noalign{\smallskip}
   	\noalign{\smallskip}
	Star & T$_\mathrm{eff}$ & V & $\sigma$ & $n$ & Cr & $\sigma$ & $n$ & Mn & $\sigma$ & $n$ & Co & $\sigma$ & $n$ & Ni & $\sigma$ & $n$ \\
	\noalign{\smallskip}
	\noalign{\smallskip}
	\hline
	\noalign{\smallskip}
	HD 142 & 6290 & 0.08 & 0.00 & 1 & 0.05 & 0.05 & 3 & -- & -- & 0 & -- & -- & 0 & 0.03 & 0.06 & 4 \\
	HD 1237 & 5555 & 0.11 & 0.07 & 3 & 0.12 & 0.05 & 6 & 0.18 & 0.10 & 2 & 0.01 & 0.03 & 3 & 0.07 & 0.04 & 10 \\
	HD 2039 & 5990 & 0.34 & 0.12 & 4 & 0.33 & 0.05 & 6 & 0.50 & 0.00 & 1 & 0.45 & 0.04 & 3 & 0.35 & 0.06 & 10 \\
	HD 4203 & 5650 & 0.40 & 0.07 & 4 & 0.36 & 0.06 & 7 & 0.54 & 0.17 & 2 & 0.54 & 0.01 & 3 & 0.45 & 0.05 & 9 \\
	HD 4208 & 5625 & $-$0.29 & 0.10 & 4 & $-$0.28 & 0.05 & 7 & $-$0.41 & 0.00 & 1 & $-$0.29 & 0.00 & 3 & $-$0.23 & 0.05 & 10 \\
	HD 6434 & 5790 & $-$0.45 & 0.07 & 3 & $-$0.55 & 0.03 & 4 & -- & -- & 0 & $-$0.51 & 0.09 & 2 & $-$0.54 & 0.05 & 10 \\
	HD 8574 & 6080 & $-$0.08 & 0.00 & 2 & $-$0.06 & 0.03 & 6 & $-$0.18 & 0.00 & 1 & $-$0.09 & 0.19 & 2 & $-$0.03 & 0.03 & 9 \\
	HD 9826 & 6120 & $-$0.05 & 0.01 & 2 & 0.03 & 0.02 & 3 & $-$0.02 & 0.00 & 1 & -- & -- & 0 & 0.02 & 0.07 & 4 \\
	HD 10697 & 5665 & 0.14 & 0.09 & 4 & 0.11 & 0.03 & 7 & 0.12 & 0.09 & 2 & 0.15 & 0.09 & 5 & 0.11 & 0.05 & 10 \\
	HD 12661 & 5715 & 0.38 & 0.07 & 5 & 0.31 & 0.06 & 7 & 0.50 & 0.16 & 2 & 0.48 & 0.05 & 4 & 0.36 & 0.07 & 10 \\
	HD 13445 & 5205 & $-$0.06 & 0.01 & 3 & $-$0.19 & 0.05 & 7 & $-$0.39 & 0.05 & 2 & $-$0.07 & 0.08 & 4 & $-$0.21 & 0.06 & 10 \\
	HD 16141 & 5805 & 0.12 & 0.02 & 4 & 0.16 & 0.05 & 7 & 0.05 & 0.09 & 2 & 0.14 & 0.06 & 4 & 0.10 & 0.05 & 10 \\
	HD 17051 & 6225 & 0.24 & 0.07 & 3 & 0.21 & 0.06 & 4 & 0.25 & 0.00 & 1 & $-$0.02 & 0.12 & 2 & 0.19 & 0.03 & 10 \\
	HD 19994 & 6175 & 0.17 & 0.09 & 3 & 0.14 & 0.01 & 3 & 0.13 & 0.00 & 1 & -- & -- & 0 & 0.20 & 0.05 & 7 \\
	HD 20367 & 6100 & 0.07 & 0.06 & 3 & 0.09 & 0.05 & 4 & 0.00 & 0.00 & 1 & $-$0.21 & 0.00 & 1 & $-$0.02 & 0.09 & 10 \\
	HD 22049 & 5135 & 0.07 & 0.11 & 4 & $-$0.09 & 0.03 & 8 & $-$0.09 & 0.11 & 2 & $-$0.09 & 0.10 & 5 & $-$0.16 & 0.04 & 10 \\
	HD 23079 & 5945 & $-$0.22 & 0.03 & 3 & $-$0.19 & 0.05 & 6 & $-$0.28 & 0.00 & 1 & $-$0.30 & 0.09 & 3 & $-$0.18 & 0.06 & 10 \\
	HD 23596 & 6125 & 0.37 & 0.11 & 5 & 0.27 & 0.04 & 7 & 0.31 & 0.00 & 1 & 0.34 & 0.09 & 3 & 0.32 & 0.07 & 10 \\
	HD 27442 & 4890 & 0.65 & 0.14 & 3 & 0.25 & 0.04 & 7 & 0.69 & 0.30 & 2 & 0.84 & 0.08 & 4 & 0.45 & 0.11 & 9 \\
	HD 28185 & 5705 & 0.31 & 0.07 & 3 & 0.24 & 0.03 & 7 & 0.42 & 0.00 & 1 & 0.38 & 0.09 & 4 & 0.31 & 0.04 & 10 \\
	HD 30177 & 5590 & 0.45 & 0.13 & 3 & 0.34 & 0.03 & 7 & 0.41 & 0.34 & 2 & 0.56 & 0.16 & 4 & 0.38 & 0.09 & 10 \\
	HD 33636 & 5990 & $-$0.09 & 0.06 & 3 & $-$0.12 & 0.05 & 4 & $-$0.28 & 0.00 & 1 & $-$0.17 & 0.10 & 2 & $-$0.12 & 0.06 & 10 \\
	HD 37124 & 5565 & $-$0.30 & 0.11 & 3 & $-$0.46 & 0.06 & 7 & $-$0.72 & 0.00 & 1 & $-$0.31 & 0.02 & 2 & $-$0.42 & 0.05 & 10 \\
	HD 38529 & 5675 & 0.40 & 0.02 & 3 & 0.34 & 0.07 & 7 & 0.33 & 0.00 & 1 & 0.55 & 0.11 & 4 & 0.40 & 0.06 & 10 \\
	HD 39091 & 5995 & 0.07 & 0.08 & 3 & 0.05 & 0.05 & 6 & 0.06 & 0.00 & 1 & 0.04 & 0.09 & 2 & 0.08 & 0.04 & 10 \\
	HD 46375 & 5315 & 0.34 & 0.04 & 3 & 0.21 & 0.05 & 7 & 0.41 & 0.23 & 2 & 0.42 & 0.14 & 5 & 0.24 & 0.03 & 9 \\
	HD 50554 & 6050 & $-$0.06 & 0.05 & 3 & $-$0.04 & 0.05 & 6 & $-$0.15 & 0.00 & 1 & $-$0.01 & 0.30 & 2 & $-$0.04 & 0.04 & 10 \\
	HD 52265 & 6098 & 0.18 & 0.07 & 3 & 0.20 & 0.04 & 5 & 0.17 & 0.00 & 1 & 0.06 & 0.01 & 2 & 0.21 & 0.04 & 10 \\
	HD 74156 & 6105 & 0.11 & 0.08 & 3 & 0.08 & 0.04 & 6 & 0.02 & 0.00 & 1 & 0.07 & 0.11 & 2 & 0.13 & 0.04 & 10 \\
	HD 75289 & 6135 & 0.25 & 0.09 & 3 & 0.19 & 0.02 & 3 & 0.15 & 0.00 & 1 & 0.12 & 0.02 & 2 & 0.21 & 0.04 & 10 \\
	HD 75732 & 5307 & 0.52 & 0.08 & 3 & 0.29 & 0.04 & 8 & 0.53 & 0.27 & 2 & 0.64 & 0.16 & 4 & 0.39 & 0.09 & 10 \\
	HD 80606 & 5570 & 0.47 & 0.19 & 5 & 0.35 & 0.07 & 7 & 0.52 & 0.11 & 2 & 0.52 & 0.14 & 5 & 0.40 & 0.06 & 9 \\
	HD 82943 & 6025 & 0.30 & 0.08 & 4 & 0.25 & 0.03 & 7 & 0.29 & 0.09 & 2 & 0.21 & 0.05 & 5 & 0.31 & 0.04 & 10 \\
	HD 83443 & 5500 & 0.46 & 0.05 & 3 & 0.32 & 0.06 & 8 & 0.64 & 0.27 & 2 & 0.56 & 0.17 & 5 & 0.42 & 0.07 & 10 \\
	HD 92788 & 5820 & 0.40 & 0.05 & 5 & 0.30 & 0.05 & 8 & 0.40 & 0.11 & 3 & 0.42 & 0.10 & 5 & 0.37 & 0.06 & 10 \\
	HD 95128 & 5925 & 0.05 & 0.05 & 3 & 0.02 & 0.05 & 7 & 0.04 & 0.07 & 2 & $-$0.02 & 0.06 & 5 & 0.04 & 0.04 & 9 \\
	HD 106252 & 5890 & $-$0.11 & 0.05 & 3 & $-$0.09 & 0.02 & 5 & $-$0.14 & 0.00 & 1 & $-$0.10 & 0.12 & 2 & $-$0.09 & 0.04 & 10 \\
	HD 108147 & 6265 & 0.18 & 0.11 & 3 & 0.13 & 0.03 & 3 & -- & -- & 0 & -- & -- & 0 & 0.10 & 0.04 & 8 \\
	HD 108874 & 5615 & 0.23 & 0.11 & 4 & 0.20 & 0.08 & 6 & 0.24 & 0.08 & 3 & 0.27 & 0.21 & 5 & 0.27 & 0.06 & 9 \\
	HD 114386 & 4875 & 0.34 & 0.14 & 4 & $-$0.04 & 0.07 & 8 & 0.11 & 0.12 & 3 & 0.16 & 0.20 & 5 & 0.02 & 0.09 & 10 \\
	HD 114729 & 5820 & $-$0.25 & 0.17 & 4 & $-$0.33 & 0.10 & 6 & $-$0.49 & 0.00 & 1 & $-$0.38 & 0.13 & 4 & $-$0.31 & 0.04 & 9 \\
	HD 114762 & 5870 & $-$0.71 & 0.06 & 3 & $-$0.87 & 0.22 & 2 & $-$1.18 & 0.00 & 1 & $-$0.87 & 0.08 & 4 & $-$0.75 & 0.04 & 9 \\
	HD 114783 & 5160 & 0.33 & 0.01 & 2 & 0.11 & 0.03 & 8 & 0.27 & 0.19 & 2 & 0.32 & 0.08 & 4 & 0.19 & 0.08 & 10 \\
	HD 117176 & 5530 & $-$0.07 & 0.07 & 4 & $-$0.11 & 0.03 & 7 & $-$0.16 & 0.00 & 1 & $-$0.08 & 0.10 & 5 & $-$0.10 & 0.03 & 9 \\
	HD 121504 & 6090 & 0.18 & 0.07 & 3 & 0.14 & 0.04 & 6 & 0.12 & 0.00 & 1 & 0.09 & 0.10 & 2 & 0.13 & 0.05 & 10 \\
	HD 128311 & 4950 & 0.25 & 0.09 & 2 & 0.03 & 0.10 & 6 & 0.16 & 0.09 & 3 & 0.21 & 0.09 & 4 & 0.05 & 0.11 & 9 \\
	HD 130322 & 5430 & 0.12 & 0.18 & 4 & 0.06 & 0.03 & 8 & 0.10 & 0.09 & 2 & 0.11 & 0.13 & 5 & 0.06 & 0.04 & 10 \\
	HD 134987 & 5780 & 0.35 & 0.07 & 4 & 0.28 & 0.03 & 8 & 0.29 & 0.26 & 3 & 0.41 & 0.13 & 5 & 0.36 & 0.04 & 10 \\
	HD 136118 & 6175 & $-$0.20 & 0.00 & 2 & $-$0.07 & 0.05 & 3 & $-$0.24 & 0.00 & 1 & -- & -- & 0 & $-$0.14 & 0.04 & 8 \\
	HD 137759 & 4750 & 0.29 & 0.09 & 3 & 0.08 & 0.10 & 8 & 0.23 & 0.02 & 2 & 0.47 & 0.21 & 5 & 0.11 & 0.15 & 9 \\
	HD 141937 & 5925 & 0.13 & 0.06 & 5 & 0.12 & 0.03 & 8 & 0.02 & 0.14 & 2 & 0.03 & 0.06 & 5 & 0.13 & 0.03 & 10 \\
	HD 143761 & 5835 & $-$0.16 & 0.04 & 3 & $-$0.28 & 0.03 & 7 & $-$0.44 & 0.00 & 1 & $-$0.26 & 0.04 & 5 & $-$0.25 & 0.05 & 10 \\
	HD 145675 & 5255 & 0.58 & 0.12 & 2 & 0.39 & 0.10 & 8 & 0.51 & 0.00 & 1 & 0.77 & 0.24 & 5 & 0.60 & 0.09 & 10 \\
	HD 147513 & 5880 & 0.03 & 0.03 & 3 & 0.03 & 0.04 & 7 & $-$0.06 & 0.00 & 1 & $-$0.09 & 0.06 & 5 & $-$0.04 & 0.03 & 10 \\
	HD 150706 & 6000 & $-$0.04 & 0.06 & 3 & $-$0.04 & 0.05 & 5 & $-$0.22 & 0.09 & 2 & $-$0.27 & 0.03 & 3 & $-$0.08 & 0.05 & 9 \\
	HD 160691 & 5820 & 0.33 & 0.08 & 5 & 0.28 & 0.04 & 7 & 0.37 & 0.07 & 2 & 0.39 & 0.09 & 5 & 0.34 & 0.04 & 10 \\
	HD 162020 & 4830 & 0.18 & 0.11 & 4 & $-$0.07 & 0.06 & 7 & 0.01 & 0.00 & 1 & 0.08 & 0.15 & 3 & $-$0.07 & 0.09 & 9 \\
	HD 168443 & 5600 & 0.10 & 0.10 & 4 & 0.02 & 0.05 & 8 & $-$0.02 & 0.11 & 2 & 0.15 & 0.09 & 5 & 0.05 & 0.05 & 10 \\
	HD 168746 & 5610 & 0.02 & 0.12 & 3 & $-$0.12 & 0.04 & 7 & $-$0.23 & 0.00 & 1 & 0.02 & 0.11 & 4 & $-$0.08 & 0.06 & 10 \\
	HD 169830 & 6300 & 0.13 & 0.12 & 3 & 0.09 & 0.02 & 6 & 0.03 & 0.00 & 1 & $-$0.09 & 0.00 & 1 & 0.10 & 0.04 & 10 \\
	HD 177830 & 4840 & 0.50 & 0.08 & 3 & 0.19 & 0.04 & 7 & 0.68 & 0.41 & 2 & 0.72 & 0.13 & 3 & 0.41 & 0.13 & 10 \\
	HD 179949 & 6235 & 0.05 & 0.00 & 2 & 0.13 & 0.03 & 3 & 0.07 & 0.00 & 1 & -- & -- & 0 & 0.14 & 0.06 & 7 \\
	HD 186427 & 5765 & 0.11 & 0.10 & 4 & 0.04 & 0.03 & 8 & 0.06 & 0.02 & 2 & 0.09 & 0.06 & 5 & 0.08 & 0.03 & 10 \\
	HD 187123 & 5855 & 0.11 & 0.03 & 3 & 0.08 & 0.03 & 8 & 0.13 & 0.04 & 2 & 0.09 & 0.06 & 5 & 0.12 & 0.03 & 10 \\
	HD 190228 & 5340 & $-$0.14 & 0.04 & 3 & $-$0.29 & 0.03 & 7 & $-$0.35 & 0.00 & 1 & $-$0.16 & 0.13 & 5 & $-$0.25 & 0.05 & 10 \\
	HD 190360 & 5590 & 0.23 & 0.09 & 4 & 0.19 & 0.05 & 6 & 0.25 & 0.17 & 2 & 0.35 & 0.15 & 5 & 0.27 & 0.03 & 8 \\
	HD 192263 & 4995 & 0.15 & 0.04 & 3 & $-$0.01 & 0.04 & 6 & 0.09 & 0.15 & 2 & 0.14 & 0.13 & 3 & 0.03 & 0.05 & 10 \\
	HD 195019 & 5840 & 0.07 & 0.03 & 4 & 0.05 & 0.02 & 6 & 0.02 & 0.00 & 1 & 0.03 & 0.03 & 4 & 0.04 & 0.04 & 10 \\
	HD 196050 & 5905 & 0.21 & 0.01 & 3 & 0.18 & 0.02 & 7 & 0.33 & 0.00 & 1 & 0.23 & 0.06 & 4 & 0.22 & 0.06 & 10 \\
	HD 202206 & 5765 & 0.31 & 0.12 & 4 & 0.27 & 0.03 & 6 & 0.47 & 0.00 & 1 & 0.45 & 0.06 & 2 & 0.33 & 0.04 & 10 \\
	HD 209458 & 6120 & 0.04 & 0.09 & 5 & $-$0.03 & 0.04 & 4 & $-$0.08 & 0.00 & 1 & $-$0.13 & 0.07 & 5 & $-$0.04 & 0.03 & 9 \\
	HD 210277 & 5570 & 0.26 & 0.00 & 3 & 0.18 & 0.05 & 8 & 0.19 & 0.13 & 2 & 0.33 & 0.12 & 5 & 0.22 & 0.05 & 10 \\
	HD 213240 & 5975 & 0.15 & 0.03 & 3 & 0.10 & 0.06 & 6 & 0.07 & 0.00 & 1 & 0.09 & 0.04 & 2 & 0.14 & 0.04 & 10 \\
	HD 216435 & 5905 & 0.13 & 0.10 & 3 & 0.11 & 0.04 & 5 & 0.15 & 0.00 & 1 & 0.11 & 0.04 & 3 & 0.18 & 0.04 & 10 \\
	HD 216437 & 5875 & 0.22 & 0.08 & 4 & 0.14 & 0.08 & 8 & 0.29 & 0.00 & 1 & 0.23 & 0.05 & 5 & 0.22 & 0.04 & 10 \\
	HD 217014 & 5805 & 0.20 & 0.06 & 5 & 0.16 & 0.01 & 6 & 0.24 & 0.00 & 1 & 0.26 & 0.01 & 3 & 0.21 & 0.03 & 10 \\
	HD 217107 & 5658 & 0.40 & 0.11 & 5 & 0.34 & 0.04 & 7 & 0.51 & 0.26 & 2 & 0.47 & 0.13 & 5 & 0.41 & 0.06 & 10 \\
	HD 222582 & 5850 & 0.06 & 0.04 & 3 & $-$0.02 & 0.04 & 6 & $-$0.04 & 0.00 & 1 & 0.08 & 0.16 & 3 & $-$0.05 & 0.12 & 8 \\
	\noalign{\smallskip}
	\hline
   	\end{tabular}
	\end{scriptsize}
\end{table*}

\begin{table*}[thbp]
	\caption{ Same as Table\,\ref{tab2} for the comparison sample (stars without giant planets). }
	\label{tab3}
	\begin{scriptsize}
   	\begin{tabular}{lcccrrcrrcrrcrrcr}	\hline
   	\noalign{\smallskip}
   	\noalign{\smallskip}
	Star & T$_\mathrm{eff}$ & $\log g$ & $\xi_{t}$ & Fe & Si & $\sigma$ & $n$ & Ca & $\sigma$ & $n$ & Sc & $\sigma$ & $n$ & Ti & $\sigma$ & $n$ \\
	\noalign{\smallskip}
	\noalign{\smallskip}
	\hline
	\noalign{\smallskip}
	HD 1581 & 5940 & 4.41 & 1.13 & $-$0.15 & $-$0.13 & 0.06 & 11 & $-$0.11 & 0.03 & 9 & $-$0.10 & 0.05 & 7 & $-$0.12 & 0.06 & 10 \\
	HD 4391 & 5955 & 4.85 & 1.22 & 0.01 & $-$0.04 & 0.06 & 11 & $-$0.02 & 0.04 & 9 & 0.07 & 0.08 & 5 & 0.06 & 0.03 & 7 \\
	HD 5133 & 5015 & 4.82 & 0.92 & $-$0.08 & $-$0.13 & 0.07 & 11 & $-$0.14 & 0.12 & 10 & 0.09 & 0.19 & 4 & 0.07 & 0.06 & 10 \\
	HD 7570 & 6135 & 4.42 & 1.46 & 0.17 & 0.19 & 0.03 & 11 & 0.14 & 0.02 & 9 & 0.17 & 0.05 & 6 & 0.12 & 0.02 & 6 \\
	HD 10360 & 5045 & 4.77 & 0.89 & $-$0.19 & $-$0.18 & 0.05 & 11 & $-$0.29 & 0.07 & 9 & $-$0.18 & 0.11 & 5 & $-$0.07 & 0.05 & 10 \\
	HD 10647 & 6130 & 4.45 & 1.31 & $-$0.03 & $-$0.05 & 0.03 & 11 & $-$0.01 & 0.04 & 10 & $-$0.08 & 0.05 & 4 & $-$0.10 & 0.02 & 4 \\
	HD 10700 & 5370 & 4.70 & 1.01 & $-$0.50 & $-$0.38 & 0.04 & 11 & $-$0.41 & 0.06 & 10 & $-$0.35 & 0.06 & 6 & $-$0.23 & 0.04 & 10 \\
	HD 14412 & 5410 & 4.70 & 1.01 & $-$0.44 & $-$0.43 & 0.04 & 11 & $-$0.44 & 0.04 & 9 & $-$0.33 & 0.04 & 4 & $-$0.35 & 0.05 & 10 \\
	HD 17925 & 5220 & 4.60 & 1.44 & 0.08 & 0.03 & 0.06 & 11 & 0.07 & 0.06 & 9 & 0.05 & 0.15 & 6 & 0.13 & 0.03 & 10 \\
	HD 20010 & 6240 & 4.27 & 2.23 & $-$0.20 & $-$0.18 & 0.05 & 11 & $-$0.26 & 0.04 & 10 & $-$0.18 & 0.09 & 5 & $-$0.16 & 0.06 & 6 \\
	HD 20766 & 5770 & 4.68 & 1.24 & $-$0.20 & $-$0.19 & 0.02 & 11 & $-$0.20 & 0.03 & 10 & $-$0.12 & 0.09 & 6 & $-$0.13 & 0.02 & 8 \\
	HD 20794 & 5465 & 4.62 & 1.04 & $-$0.36 & $-$0.19 & 0.03 & 11 & $-$0.24 & 0.05 & 10 & $-$0.13 & 0.09 & 6 & $-$0.04 & 0.05 & 8 \\
	HD 20807 & 5865 & 4.59 & 1.28 & $-$0.22 & $-$0.20 & 0.02 & 11 & $-$0.24 & 0.02 & 10 & $-$0.13 & 0.07 & 6 & $-$0.22 & 0.04 & 8 \\
	HD 23249 & 5135 & 4.00 & 1.12 & 0.17 & 0.18 & 0.06 & 11 & 0.10 & 0.07 & 10 & 0.20 & 0.08 & 6 & 0.25 & 0.11 & 10 \\
	HD 23356 & 5035 & 4.73 & 0.96 & $-$0.05 & $-$0.06 & 0.05 & 11 & $-$0.10 & 0.08 & 9 & 0.08 & 0.18 & 4 & 0.10 & 0.07 & 10 \\
	HD 23484 & 5230 & 4.62 & 1.13 & 0.10 & 0.07 & 0.04 & 11 & 0.06 & 0.08 & 9 & 0.06 & 0.09 & 5 & 0.15 & 0.05 & 10 \\
	HD 26965A & 5185 & 4.73 & 0.75 & $-$0.26 & $-$0.11 & 0.04 & 11 & $-$0.21 & 0.08 & 9 & $-$0.07 & 0.06 & 5 & 0.08 & 0.04 & 9 \\
	HD 30495 & 5880 & 4.67 & 1.29 & 0.03 & 0.00 & 0.04 & 11 & 0.02 & 0.03 & 10 & 0.02 & 0.06 & 6 & 0.03 & 0.06 & 9 \\
	HD 36435 & 5510 & 4.78 & 1.15 & 0.03 & $-$0.02 & 0.03 & 11 & $-$0.02 & 0.03 & 9 & 0.04 & 0.10 & 6 & 0.03 & 0.03 & 10 \\
	HD 38858 & 5750 & 4.56 & 1.22 & $-$0.22 & $-$0.20 & 0.03 & 11 & $-$0.21 & 0.03 & 10 & $-$0.18 & 0.07 & 5 & $-$0.18 & 0.04 & 10 \\
	HD 40307 & 4925 & 4.57 & 0.79 & $-$0.25 & $-$0.27 & 0.06 & 9 & $-$0.18 & 0.08 & 9 & $-$0.24 & 0.10 & 5 & 0.03 & 0.07 & 10 \\
	HD 43162 & 5630 & 4.57 & 1.36 & $-$0.02 & $-$0.03 & 0.03 & 11 & $-$0.02 & 0.06 & 9 & $-$0.02 & 0.10 & 5 & $-$0.03 & 0.04 & 8 \\
	HD 43834 & 5620 & 4.56 & 1.10 & 0.12 & 0.14 & 0.03 & 11 & 0.07 & 0.06 & 10 & 0.18 & 0.10 & 7 & 0.13 & 0.05 & 10 \\
	HD 50281A & 4790 & 4.75 & 0.85 & 0.07 & $-$0.04 & 0.04 & 8 & $-$0.03 & 0.11 & 10 & 0.11 & 0.16 & 4 & 0.15 & 0.07 & 9 \\
	HD 53705 & 5810 & 4.40 & 1.18 & $-$0.19 & $-$0.14 & 0.03 & 11 & $-$0.17 & 0.03 & 10 & $-$0.08 & 0.07 & 6 & $-$0.13 & 0.03 & 8 \\
	HD 53706 & 5315 & 4.50 & 0.90 & $-$0.22 & $-$0.16 & 0.05 & 11 & $-$0.18 & 0.06 & 9 & $-$0.13 & 0.12 & 6 & $-$0.03 & 0.03 & 10 \\
	HD 65907A & 5940 & 4.56 & 1.19 & $-$0.29 & $-$0.12 & 0.02 & 11 & $-$0.16 & 0.03 & 10 & $-$0.08 & 0.08 & 7 & $-$0.06 & 0.07 & 9 \\
	HD 69830 & 5455 & 4.56 & 0.98 & 0.00 & 0.00 & 0.03 & 11 & $-$0.02 & 0.05 & 9 & 0.06 & 0.10 & 6 & 0.08 & 0.05 & 10 \\
	HD 72673 & 5290 & 4.68 & 0.81 & $-$0.33 & $-$0.33 & 0.04 & 11 & $-$0.32 & 0.05 & 9 & $-$0.23 & 0.09 & 5 & $-$0.18 & 0.03 & 10 \\
	HD 74576 & 5080 & 4.86 & 1.20 & 0.04 & $-$0.01 & 0.05 & 11 & $-$0.06 & 0.08 & 9 & 0.08 & 0.16 & 6 & 0.09 & 0.06 & 10 \\
	HD 76151 & 5825 & 4.62 & 1.08 & 0.15 & 0.14 & 0.02 & 11 & 0.11 & 0.04 & 10 & 0.15 & 0.03 & 4 & 0.14 & 0.03 & 10 \\
	HD 84117 & 6140 & 4.35 & 1.38 & $-$0.04 & $-$0.03 & 0.03 & 11 & $-$0.06 & 0.05 & 10 & $-$0.05 & 0.07 & 4 & $-$0.06 & 0.07 & 5 \\
	HD 189567 & 5750 & 4.57 & 1.21 & $-$0.23 & $-$0.18 & 0.02 & 11 & $-$0.22 & 0.04 & 10 & $-$0.14 & 0.07 & 7 & $-$0.13 & 0.04 & 9 \\
	HD 191408A & 5025 & 4.62 & 0.74 & $-$0.51 & $-$0.40 & 0.04 & 10 & $-$0.33 & 0.09 & 9 & $-$0.38 & 0.09 & 4 & $-$0.04 & 0.06 & 10 \\
	HD 192310 & 5125 & 4.63 & 0.88 & 0.05 & 0.09 & 0.06 & 11 & $-$0.01 & 0.08 & 9 & 0.18 & 0.14 & 7 & 0.17 & 0.07 & 10 \\
	HD 196761 & 5460 & 4.62 & 1.00 & 0.27 & $-$0.24 & 0.03 & 11 & $-$0.28 & 0.02 & 9 & $-$0.21 & 0.10 & 5 & $-$0.19 & 0.06 & 10 \\
	HD 207129 & 5910 & 4.53 & 1.21 & $-$0.01 & 0.00 & 0.03 & 11 & $-$0.01 & 0.03 & 10 & 0.07 & 0.08 & 6 & $-$0.02 & 0.07 & 9 \\
	HD 209100 & 4700 & 4.68 & 0.60 & 0.01 & $-$0.05 & 0.06 & 9 & $-$0.19 & 0.12 & 9 & 0.05 & 0.16 & 6 & 0.10 & 0.11 & 10 \\
	HD 211415 & 5925 & 4.65 & 1.27 & $-$0.16 & $-$0.17 & 0.04 & 11 & $-$0.21 & 0.03 & 9 & $-$0.10 & 0.10 & 6 & $-$0.13 & 0.03 & 8 \\
	HD 216803 & 4647 & 4.88 & 0.90 & 0.07 & $-$0.01 & 0.09 & 8 & $-$0.20 & 0.09 & 9 & 0.18 & 0.02 & 2 & 0.05 & 0.11 & 10 \\
	HD 222237 & 4770 & 4.79 & 0.35 & $-$0.22 & $-$0.20 & 0.04 & 9 & $-$0.29 & 0.11 & 9 & $-$0.05 & 0.09 & 3 & 0.10 & 0.10 & 9 \\
	HD 222335 & 5310 & 4.64 & 0.97 & $-$0.10 & $-$0.13 & 0.06 & 11 & $-$0.12 & 0.06 & 9 & $-$0.17 & 0.12 & 6 & $-$0.02 & 0.05 & 10 \\
	\noalign{\smallskip}
	\hline
   	\end{tabular}
	\end{scriptsize}
\end{table*}

\begin{table*}[t]
	\caption{ Same as Table\,\ref{tab2b} for the comparison sample (stars without giant planets).}
	\label{tab3b}
	\begin{scriptsize}
   	\begin{tabular}{lcrcrrcrrcrrcrrcr}	\hline
   	\noalign{\smallskip}
   	\noalign{\smallskip}
	Star & T$_\mathrm{eff}$ & V & $\sigma$ & $n$ & Cr & $\sigma$ & $n$ & Mn & $\sigma$ & $n$ & Co & $\sigma$ & $n$ & Ni & $\sigma$ & $n$ \\
	\noalign{\smallskip}
	\noalign{\smallskip}
	\hline
	\noalign{\smallskip}
	HD 1581 & 5940 & $-$0.20 & 0.11 & 4 & $-$0.21 & 0.02 & 6 & $-$0.27 & 0.12 & 2 & $-$0.20 & 0.10 & 3 & $-$0.22 & 0.05 & 10 \\
	HD 4391 & 5955 & 0.05 & 0.02 & 3 & $-$0.07 & 0.09 & 6 & $-$0.23 & 0.08 & 2 & $-$0.13 & 0.00 & 1 & $-$0.05 & 0.06 & 10 \\
	HD 5133 & 5015 & 0.04 & 0.23 & 5 & $-$0.13 & 0.04 & 7 & $-$0.08 & 0.04 & 2 & 0.01 & 0.13 & 3 & $-$0.16 & 0.08 & 9 \\
	HD 7570 & 6135 & 0.16 & 0.10 & 3 & 0.13 & 0.03 & 6 & 0.14 & 0.00 & 1 & 0.06 & 0.06 & 4 & 0.14 & 0.06 & 10 \\
	HD 10360 & 5045 & $-$0.06 & 0.04 & 3 & $-$0.24 & 0.03 & 7 & $-$0.27 & 0.06 & 2 & $-$0.10 & 0.11 & 5 & $-$0.23 & 0.04 & 10 \\
	HD 10647 & 6130 & $-$0.03 & 0.10 & 2 & $-$0.09 & 0.05 & 6 & $-$0.21 & 0.00 & 1 & $-$0.38 & 0.00 & 1 & $-$0.14 & 0.04 & 10 \\
	HD 10700 & 5370 & $-$0.31 & 0.03 & 3 & $-$0.51 & 0.03 & 6 & $-$0.73 & 0.00 & 1 & $-$0.40 & 0.05 & 1 & $-$0.50 & 0.04 & 10 \\
	HD 14412 & 5410 & $-$0.35 & 0.04 & 3 & $-$0.49 & 0.05 & 7 & $-$0.61 & 0.00 & 1 & $-$0.45 & 0.02 & 3 & $-$0.49 & 0.04 & 10 \\
	HD 17925 & 5220 & 0.14 & 0.14 & 5 & 0.11 & 0.05 & 7 & 0.07 & 0.08 & 2 & 0.08 & 0.08 & 4 & 0.01 & 0.07 & 10 \\
	HD 20010 & 6240 & $-$0.11 & 0.12 & 3 & $-$0.29 & 0.09 & 6 & $-$0.46 & 0.00 & 1 & $-$0.45 & 0.00 & 1 & $-$0.23 & 0.07 & 10 \\
	HD 20766 & 5770 & $-$0.18 & 0.03 & 4 & $-$0.24 & 0.06 & 7 & $-$0.29 & 0.00 & 1 & $-$0.28 & 0.05 & 3 & $-$0.23 & 0.03 & 10 \\
	HD 20794 & 5465 & $-$0.25 & 0.15 & 5 & $-$0.38 & 0.04 & 7 & $-$0.57 & 0.00 & 1 & $-$0.21 & 0.05 & 5 & $-$0.35 & 0.02 & 10 \\
	HD 20807 & 5865 & $-$0.22 & 0.04 & 4 & $-$0.26 & 0.03 & 7 & $-$0.32 & 0.06 & 2 & $-$0.31 & 0.03 & 3 & $-$0.26 & 0.03 & 10 \\
	HD 23249 & 5135 & 0.34 & 0.05 & 3 & 0.07 & 0.04 & 6 & 0.28 & 0.26 & 2 & 0.36 & 0.08 & 3 & 0.21 & 0.08 & 10 \\
	HD 23356 & 5035 & 0.15 & 0.07 & 3 & $-$0.08 & 0.03 & 7 & $-$0.02 & 0.01 & 2 & 0.09 & 0.10 & 4 & $-$0.07 & 0.06 & 10 \\
	HD 23484 & 5230 & 0.24 & 0.10 & 4 & 0.09 & 0.03 & 7 & 0.14 & 0.12 & 2 & 0.16 & 0.09 & 4 & 0.05 & 0.05 & 10 \\
	HD 26965A & 5185 & 0.01 & 0.06 & 3 & $-$0.26 & 0.02 & 7 & $-$0.36 & 0.05 & 2 & $-$0.02 & 0.03 & 4 & $-$0.22 & 0.04 & 10 \\
	HD 30495 & 5880 & 0.01 & 0.05 & 3 & 0.05 & 0.03 & 7 & $-$0.06 & 0.00 & 1 & $-$0.04 & 0.09 & 4 & $-$0.03 & 0.04 & 10 \\
	HD 36435 & 5510 & 0.05 & 0.06 & 4 & 0.03 & 0.02 & 7 & $-$0.04 & 0.02 & 2 & $-$0.06 & 0.01 & 3 & $-$0.05 & 0.02 & 10 \\
	HD 38858 & 5750 & $-$0.25 & 0.04 & 3 & $-$0.21 & 0.04 & 7 & $-$0.36 & 0.00 & 1 & $-$0.20 & 0.03 & 3 & $-$0.25 & 0.04 & 10 \\
	HD 40307 & 4925 & 0.05 & 0.06 & 3 & $-$0.25 & 0.03 & 7 & $-$0.29 & 0.00 & 1 & $-$0.13 & 0.14 & 4 & $-$0.30 & 0.07 & 10 \\
	HD 43162 & 5630 & $-$0.05 & 0.03 & 3 & 0.00 & 0.05 & 6 & $-$0.08 & 0.00 & 1 & $-$0.16 & 0.07 & 2 & $-$0.11 & 0.06 & 10 \\
	HD 43834 & 5620 & 0.13 & 0.12 & 4 & 0.10 & 0.02 & 7 & 0.19 & 0.10 & 2 & 0.21 & 0.08 & 5 & 0.14 & 0.04 & 10 \\
	HD 50281A & 4790 & 0.30 & 0.11 & 4 & 0.00 & 0.09 & 8 & 0.09 & 0.03 & 2 & 0.21 & 0.12 & 4 & $-$0.03 & 0.04 & 9 \\
	HD 53705 & 5810 & $-$0.15 & 0.06 & 4 & $-$0.26 & 0.04 & 7 & $-$0.41 & 0.00 & 1 & $-$0.17 & 0.10 & 3 & $-$0.23 & 0.02 & 10 \\
	HD 53706 & 5315 & $-$0.05 & 0.12 & 4 & $-$0.23 & 0.01 & 7 & $-$0.29 & 0.00 & 1 & $-$0.09 & 0.09 & 5 & $-$0.22 & 0.05 & 10 \\
	HD 65907A & 5940 & $-$0.15 & 0.09 & 5 & $-$0.35 & 0.03 & 6 & $-$0.58 & 0.00 & 1 & $-$0.26 & 0.07 & 3 & $-$0.30 & 0.04 & 10 \\
	HD 69830 & 5455 & 0.12 & 0.10 & 4 & $-$0.02 & 0.03 & 7 & $-$0.01 & 0.03 & 2 & 0.05 & 0.11 & 5 & 0.00 & 0.03 & 10 \\
	HD 72673 & 5290 & $-$0.15 & 0.08 & 3 & $-$0.36 & 0.03 & 7 & $-$0.51 & 0.06 & 2 & $-$0.24 & 0.09 & 5 & $-$0.35 & 0.03 & 10 \\
	HD 74576 & 5080 & 0.16 & 0.10 & 4 & $-$0.01 & 0.05 & 7 & 0.01 & 0.05 & 2 & 0.11 & 0.01 & 3 & $-$0.03 & 0.07 & 10 \\
	HD 76151 & 5825 & 0.17 & 0.03 & 4 & 0.18 & 0.05 & 7 & 0.16 & 0.03 & 2 & 0.16 & 0.03 & 4 & 0.15 & 0.02 & 10 \\
	HD 84117 & 6140 & $-$0.10 & 0.11 & 3 & $-$0.08 & 0.05 & 5 & $-$0.18 & 0.00 & 1 & $-$0.29 & 0.04 & 2 & $-$0.08 & 0.06 & 10 \\
	HD 189567 & 5750 & $-$0.25 & 0.04 & 4 & $-$0.26 & 0.04 & 7 & $-$0.37 & 0.00 & 1 & $-$0.22 & 0.02 & 3 & $-$0.25 & 0.04 & 10 \\
	HD 191408A & 5025 & $-$0.13 & 0.04 & 3 & $-$0.46 & 0.03 & 7 & $-$0.58 & 0.12 & 2 & $-$0.27 & 0.08 & 5 & $-$0.50 & 0.04 & 10 \\
	HD 192310 & 5125 & 0.25 & 0.07 & 3 & 0.03 & 0.04 & 7 & 0.16 & 0.12 & 2 & 0.23 & 0.14 & 5 & 0.12 & 0.07 & 10 \\
	HD 196761 & 5460 & $-$0.25 & 0.09 & 3 & $-$0.26 & 0.04 & 7 & $-$0.34 & 0.02 & 2 & $-$0.24 & 0.08 & 5 & $-$0.29 & 0.02 & 10 \\
	HD 207129 & 5910 & $-$0.04 & 0.05 & 4 & $-$0.05 & 0.06 & 7 & $-$0.11 & 0.07 & 2 & $-$0.10 & 0.09 & 3 & $-$0.06 & 0.03 & 10 \\
	HD 209100 & 4700 & 0.22 & 0.11 & 4 & $-$0.10 & 0.07 & 8 & $-$0.02 & 0.07 & 2 & 0.20 & 0.14 & 4 & $-$0.06 & 0.09 & 10 \\
	HD 211415 & 5925 & $-$0.18 & 0.06 & 4 & $-$0.21 & 0.06 & 7 & $-$0.38 & 0.00 & 1 & $-$0.22 & 0.00 & 1 & $-$0.19 & 0.02 & 10 \\
	HD 216803 & 4647 & 0.17 & 0.12 & 4 & $-$0.10 & 0.08 & 8 & 0.03 & 0.03 & 2 & 0.14 & 0.05 & 3 & $-$0.04 & 0.07 & 9 \\
	HD 222237 & 4770 & 0.16 & 0.09 & 4 & $-$0.31 & 0.08 & 8 & $-$0.35 & 0.02 & 2 & 0.04 & 0.16 & 3 & $-$0.22 & 0.10 & 10 \\
	HD 222335 & 5310 & $-$0.07 & 0.03 & 3 & $-$0.10 & 0.04 & 7 & $-$0.17 & 0.01 & 2 & $-$0.06 & 0.10 & 4 & $-$0.17 & 0.04 & 10 \\
	\noalign{\smallskip}
	\hline
   	\end{tabular}
	\end{scriptsize}
\end{table*}

Non-LTE effects, that are not taken into account in our analysis, and the assumption of
plane-parallel model atmospheres, are also sources of errors. These are discussed in more
detail in the Appendix. Overall, these produce errors that are of the same order of magnitude 
(or lower) as the errors in the analysis, and are thus more or less negligible 
\citep[see also discussions in e.g. ][]{Edv93,Fel98,The99}.

\section{Comparing the samples}
\label{sec:comparing}

The major goal of this work is to check for any significant differences between the 
planet-host star sample and the sample of stars without any known giant planets, concerning
metals other than iron. There are already a few studies in the literature on this subject. \citet[][]{San00} 
compared the abundances of planet hosts and non-hosts for several elements (including C and O),
and found no statistically significant differences. \citet{Gon00} and \citet{Gon01}
have discussed some possible anomalies concerning Na, Mg and Al, in the sense that the [X/Fe] abundance
ratios for the planet-host stars seemed to be slightly lower than those found for field dwarfs. \citet[][]{Zha02} have further discussed a possible anomaly for Mg, but in contrast to \citet{Gon01},
they observed an enrichment of Mg in planet-host stars. On the other hand, \citet[][]{Sad02} found no
special (general) trends concerning any of the elements discussed above. 
Finally, both \citet[][]{Smi01} and \citet[][]{Sad02} compared the condensation-temperature 
slopes (computed as the slope of the points in the T$_c$ vs. [X/H] plane) for stars with and without
planets.
They found no significant differences, although a few particular planet-host stars deviated from the 
main trend.

This confusing situation is due, in part, to the fact that all the studies until now have been based
upon comparisons of non-uniform sets of data. Frequently, different line lists and model atmospheres 
were used in the chemical analysis for the two comparing samples. 
The data that we are presenting here gives us the possibility to solve this problem (at 
least for the elements studied). Let us then see what we find. 

\subsection{The [X/H] distributions}

\begin{figure}[t]
\psfig{width=\hsize,file=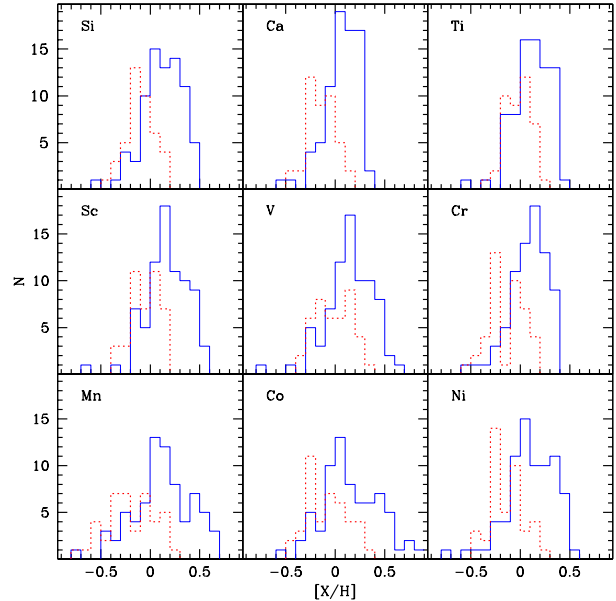}
\caption[]{[X/H] distributions of planet-host stars (solid lines) and for our comparison sample (dotted lines). }
\label{fig1}
\end{figure}

In Fig.\,\ref{fig1}, we provide the distributions
of the [X/H] for planet hosts and non-hosts.
These histograms, which are similar to the ones presented for iron in \citet[][]{San02}, indicate that 
the excess metallicity observed for planet hosts is, as expected, clearly widespread,
and is not unique to iron.

An interesting feature of these histograms is that they show that the star-with-planet sample
is usually not symmetrical. As observed for iron \citep[][]{San01,San02,Rei02}, the distributions
for the various elements (this is particularly evident for e.g. Ca, Ti and Cr) seem to be an increasing 
function of [X/H] up to a given value where the distribution falls abruptly; possible interpretations
for this are discussed in e.g. \citet[][]{San01,San02}. 

On the other hand, for a few elements (e.g. Co, and Mn), the star-with-planet 
distributions appear to be slightly bimodal. The significance and possible implications of this are not clear. It is difficult to conceive that
some of the planet hosts had been enriched only in these particular elements (producing 
the bimodal distributions). If stellar ``pollution'' were involved, we would not expect 
large differences between all the elements studied here since their condensation temperatures 
are not very different \citep[][]{Was85}. {This feature may be
related to the lack of stars in our samples with [Fe/H] around $+$0.3.}

\begin{table}[t] \centering
        \caption{ Average abundance values $<$[X/H]$>$ for stars with planets and for our 
comparison sample. Also listed are the rms around the mean and the difference between the two samples
for each element. The number of stars is 77 and 42 for the two samples mentioned, respectively. }
        \label{TabMean}
        \begin{tabular}{lccccc}  \hline
        \noalign{\smallskip}
        \noalign{\smallskip}
                &       \multicolumn{2}{c}{Comparison sample} & \multicolumn{2}{c}{Planet hosts}   &    \\
        Species &       $<$[X/H]$>$ & rms & $<$[X/H]$>$ & rms & Difference   \\
        \noalign{\smallskip}
        \noalign{\smallskip}
        \hline
        \noalign{\smallskip}
        Si      &       $-0.10$ & 0.15 & 0.11 & 0.20 &0.21\\
        Ca      &       $-0.13$ & 0.14 & 0.07 & 0.17 &0.20\\
        Ti      &       $-0.01$ & 0.13 & 0.12 & 0.18 &0.13\\
        Sc      &       $-0.04$ & 0.15 & 0.17 & 0.21 &0.21\\
        V       &       $-0.01$ & 0.18 & 0.14 & 0.24 &0.15\\
        Cr      &       $-0.14$ & 0.18 & 0.07 & 0.22 &0.21\\
        Mn      &       $-0.20$ & 0.25 & 0.10 & 0.33 &0.30\\
        Co      &       $-0.08$ & 0.20 & 0.14 & 0.31 &0.22\\
        Ni      &       $-0.14$ & 0.18 & 0.10 & 0.24 &0.24\\
        \noalign{\smallskip}
        \hline
        \end{tabular}
\end{table}

In Table\,\ref{TabMean}, we list the average values of [X/H] for each element in the
two distributions, as well as the rms dispersions and the difference between the
average [X/H] for stars with and without planetary-mass companions. The differences vary from
0.13 (for Ti) to 0.30 (for Mn). Given the usually high dispersions around the mean values, 
these discrepancies are not very significant. In any case, they probably reflect the 
``normal'' chemical evolution of the galaxy (see Sect.\ref{sec:galactic}).

\subsection{Comparing in the [X/Fe] vs. [Fe/H] plane}

\begin{figure*}[t]
\psfig{width=\hsize,file=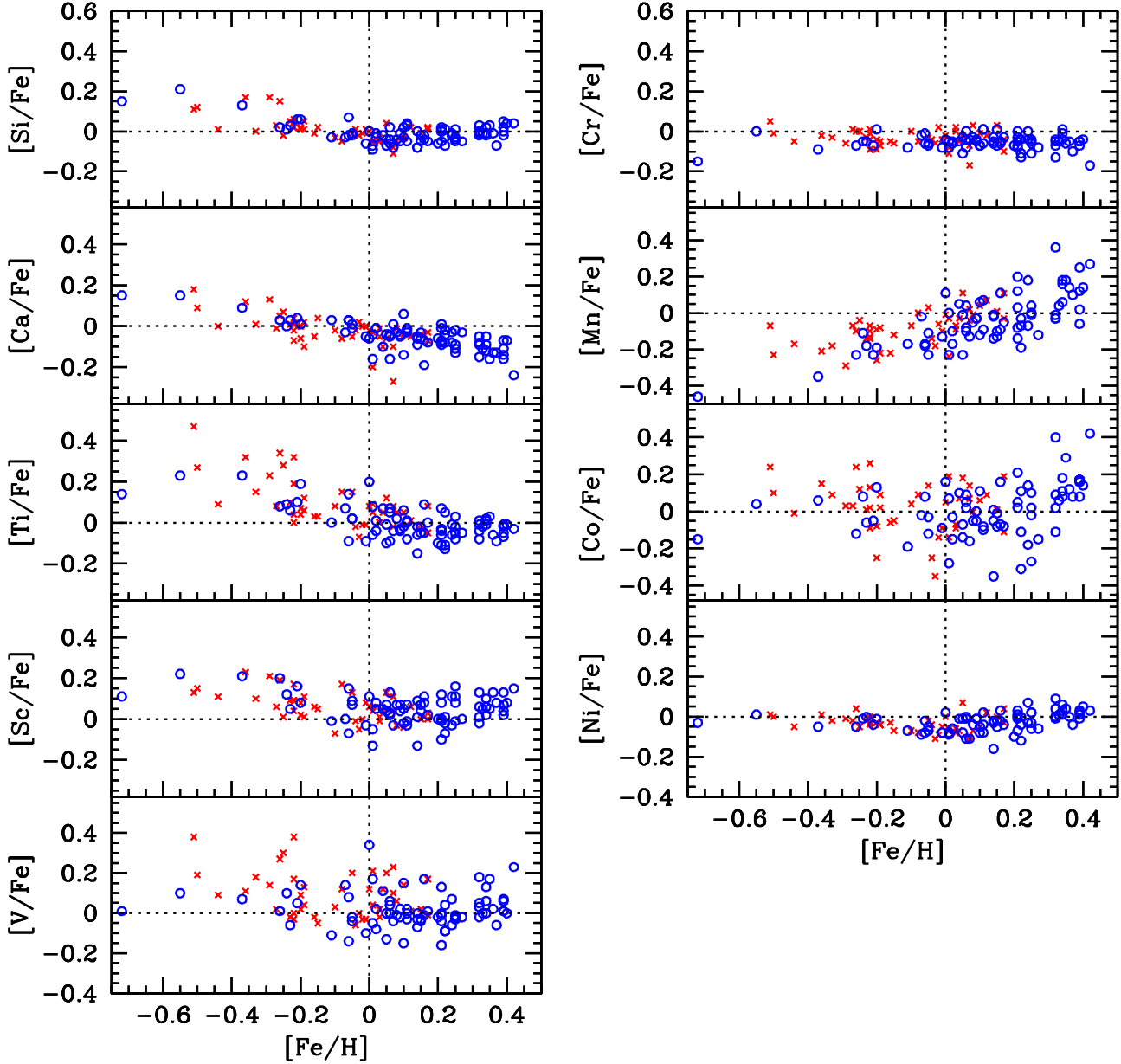}
\caption[]{[X/Fe] vs. [Fe/H] plots for the 9 elements studied in this paper.
The crosses represent the comparison-sample stars, while the open circles
denote the planet-host stars. The dotted lines represent the position of the Sun. }
\label{fig2}
\end{figure*}

Fig.\,\ref{fig2} presents the abundance ratios [X/Fe] of all 
elements as functions of [Fe/H] for both samples discussed above. Overall, 
the abundance trends for the stars with planets are nearly indistinguishable from those of
the field stars. The only conspicuous difference is the higher average iron content 
of the planet-host sample. The abundance distributions of stars with giant planets are high [Fe/H] 
extensions to the curves traced by the field dwarfs without planets (no discontinuity is seen), 
and in the regions of overlap, 
we do not find any clearly significant difference between samples. 

\begin{figure}[t]
\psfig{width=\hsize,file=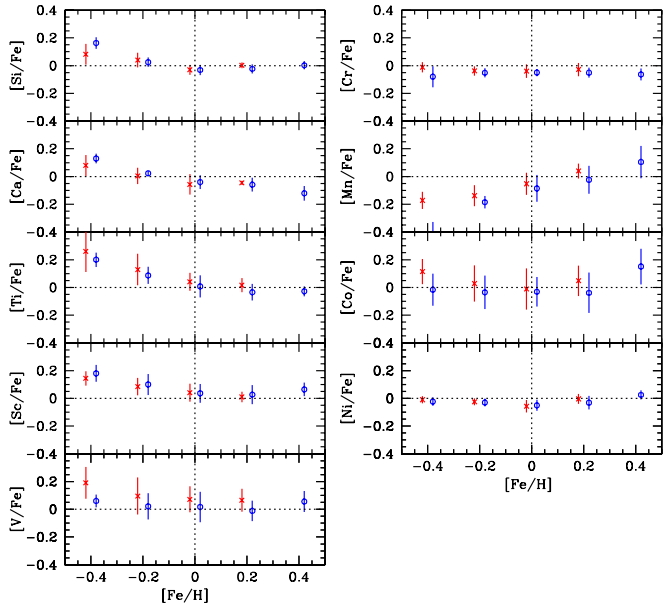}
\caption[]{Same as Fig.\,\ref{fig2}, but now using binned average values.
For both samples the bins are centered at [Fe/H]=$-$0.4, $-$0.2, 0.0, 0.2, and 0.4\,dex,
and are 0.2\,dex wide.
The crosses represent the comparison-sample stars, while the circles
denote the planet-host stars. The error bars depict the rms around the mean value.}
\label{fig3}
\end{figure}

In Fig.\,\ref{fig3}, we present the same kind of plots, but with binned average
values. For both samples, the bins are centered at [Fe/H]=$-$0.4, $-$0.2, 0.0, 0.2, and 0.4\,dex,
and are 0.2\,dex wide. These plots show that for most elements, the two groups of stars
seem to behave in the same manner.
Overall, V and Mn, and to a lesser extent Ti and Co, are alone in featuring somewhat distinguishing and systematic traits 
between the two samples. Field stars are consistently more abundant in vanadium than the planet hosts 
(up to 0.15 dex apart). \citet{Sad02} proposed a similar overabundance for [Fe/H]$\ge$0.0 dex, except that in their case, the planet-bearing stars were the ones that were vanadium-enriched. Strong
star-to-star scatter per metallicity obscures vanadium trends and could be responsible for
disagreements between the two groups (see also discussion in the next section). The same is true
for Co. 
Manganese is somewhat more prevalent in the field sample but the fact that a 
maximum of three spectral lines are available casts some doubt on this assessment. As for Ti, 
the differences found are small. 

{All these trends might be related to the NLTE effects described
in the Appendix\footnote{These 4 elements seem to suffer the
strongest NLTE effects; furthermore, 
in average, planet hosts have higher T$_{\mathrm{eff}}$ than stars
without planets by about 250K.}.} In any case, these dissimilarities are subtle, and may very well be negligible, but they are still intriguing 
enough to merit renewed tests using comparable techniques and parameters.

It is important to note that these possible differences, if confirmed, are probably not
indicators of differential accretion of material by the star, for example, since
the condensation temperatures of these elements are all in a short range of merely 
$\sim$300\,K \citep[e.g.][]{Was85} (they are all refractory). The reason would then 
probably have to do with the source of the metals examined. Whether the general trends (see discussion in 
Sect.\ref{sec:galactic}) are so that planets might form more easily around stars with
specific metallicity ratios is not excluded. Once again, this does not seem very likely
given the nature of the elements analyzed here. 

Unfortunately, this comparison is somewhat limited since the two stellar
samples overlap in a small region of [Fe/H]. In particular, the (high-)metallicity region
where most of the planet hosts are located does not contain many comparison stars. Thus, we can not completely exclude the existence of important differences
for these objects.  

\section{Galactic trends}
\label{sec:galactic}

Although the main goal of this work is to compare the elemental distributions
for stars with and without giant-planetary companions, a clear byproduct of this
study is the chance to increase our knowledge of galactic 
chemical evolution trends. This is especially true for the high [Fe/H] region,
for which the number of detailed studies is still limited. 

In this section, we will therefore make a brief comparison
of the results we have obtained with those presented in the major studies in the literature 
regarding this subject\footnote{This comparison might be seen (also) as a test
for the reliability of our analysis.}. Given the small (and probably insignificant) differences 
argued above between planet hosts and non-hosts, we will consider our sample as a whole 
in the rest of the analysis. We will also keep the discussion brief, leaving
the interpretation of the galactic chemical evolution to a future paper.

Of course, some of the trends discussed below (in particular for 
the more metal-rich stars) may eventually be seen as signatures of the presence
of planets, since these stars are (in our sample) all planet hosts. {For example, differences
between the observed trends and those published in the literature could
reflect the presence of planets, and would thus be of great importance.} However,
the trends are probably (and easily) best interpreted as simple byproducts of galactic evolution, and
their relation to the presence of a planet are probably coincidental (besides the
fact, of course, that these are the more globally metal-rich stars). 
Furthermore, given that there are no metal-rich stars that lack planets in the
current sample,
it is not possible to compare the two groups of stars in these high-[Fe/H] regions
of the plots\footnote{But as noted in the last section, there are no discontinuities
between the two samples.}. {Again, this means that we cannot exclude 
that the observed trends for the high-[Fe/H] stars are due to some
kind of planetary induced chemical variation, or that the abundance
ratios for these stars are themselves influencing the efficiency of planetary 
formation.} 

\subsection{Silicon}

From Figs.\,\ref{fig2} and \ref{fig3}, we can see that [Si/Fe] is inversely related to [Fe/H]
for $-$0.6$<$[Fe/H]$<$$-$0.2\,dex, and then appears to level off at $-$0.2$<$[Fe/H]$<$0.0\,dex.
Similar results were also found by \citet[][]{Edv93} (hereafter EAGLNT) and \citet[][]{Che00} (hereafter, C00).
As these authors have found, the dispersion in the abundance ratio [Si/Fe] is
very small in the metallicity range studied here.

For metallicities above solar, our data shows a plateau in the [Si/Fe] 
vs. [Fe/H] relation. Again, this result is compatible with the one found by both EAGLNT and C00.
However, we also see a hint of an increase in the [Si/Fe] vs. [Fe/H] trend for [Fe/H]$>$0.3 (and maybe 
before this value).
This upturn was already slightly noticeable in the data of EAGLNT, but given the low upper limit in [Fe/H] of the
objects analyzed by these authors, no serious conclusion was possible.
\citet[][]{Fel98} (hereafter FG98) analyzed stars with [Fe/H] up to $\sim$0.4\,dex,
but they found a much larger scatter in the abundances; their results do not 
permit an investigation of the behavior of [Si/Fe] for these high metallicities.
We note, however, that the data for planet-host stars of \citet[][]{Gon01} (see their Fig.\,10)
does not support the presence of such a trend.

\subsection{Calcium}

For [Fe/H] lower than solar, the behavior of [Ca/Fe] is very similar to the one described above for 
Si. Contrary to this latter element, the [Ca/Fe] ratio seems to decrease almost 
continuously for the entire metallicity range studied in this paper (see Figs.\,\ref{fig2} and \ref{fig3}).
A glance at the figures suggests the presence of a plateau in the region $-$0.2$<$[Fe/H]$<$0.2\,dex,
immediately followed, for higher metallicities, by a clear drop-off. While the former trends have been
found both by FG98 and EAGLNT (also visible in the plots of \citet[][]{Sad02}), this latter downturn was not clearly visible in the 
data presented by these authors, probably because their data never ventured above $\sim$0.4 
and 0.3\,dex, respectively. This downturn was also seen in the data of \citet[][]{Gon01}.

Once again, the dispersion of our abundance is very small (except for abundances around [Fe/H]=0.0-0.2\,dex for 
which a few stragglers are present: HD\,209100 with [Ca/Fe]=$-$0.20 and HD\,216803 with [Ca/Fe]=$-$0.27\,dex,
both very cool dwarfs). This fact gives us confidence for the reliability of the trends discussed above.
The scatter for these two objects was first explained by the existence of NLTE effects (see discussion
in FG98). However, \citet[][]{Tho00} has shown that this was not the main reason for the
observed discrepancy. Furthermore, \citet[][]{ThoFel00} demonstrated that the use of temperatures
computed from a strict excitation equilibrium (as used here) eliminates most NLTE effects.
As also discussed in the Appendix, there still seems to be a systematic
decline of the [Ca/Fe] abundance ratios with decreasing temperature for the most metal-rich stars. 
We note, though, that this effect cannot be responsible for the downturn of [Ca/Fe] seen for the higher [Fe/H] stars since the stars that occupy this region of the plot ([Fe/H]$>$0) present all kinds of effective 
temperatures.

\subsection{Titanium}

Established by FG98, EAGLNT, and C00, [Ti/Fe] seems to follow a continuous 
decline with [Fe/H]. The [Ti/Fe] distribution has a more precipitous slope than for [Si/Fe] and [Ca/Fe],
but it also carries a wider dispersion (see Fig.\,\ref{fig2}). Judging from Figs.\,\ref{fig2} and \ref{fig3},
titanium drops until $-0.2 < [$Fe$/$H$] < 0.0$ dex and then settles. This supports C00
and is unlike the continuing downward trend proposed by EAGLNT. 

The results of FG98, C00, EAGLNT, and \citet[][]{Gon01} have all reproduced the pronounced 
scatter in titanium abundances seen here. The scatter can probably be attributed to NLTE effects 
over things like line-blending or real ``physical'' galactic evolution effects. As mentioned in the 
Appendix, there is a large dependence of the [Ti/Fe] as a function of T$_{\mathrm{eff}}$, 
which is likely behind the large scatter. {Given that the stars with [Fe/H] below $-$0.2
are, in average, cooler (by about 200\,K) than the rest of the sample\footnote{We stress that
this trend is only slightly significant for this metallicity regime. Stars
with [Fe/H] between -0.1 and 0.1 have, for example, the ``same'' average T$_{\mathrm{eff}}$ as the
objects with [Fe/H] above 0.3.}, it is possible
that part (but very unlikely all) of the decreasing trend discussed above is due to NLTE effects.}.

\subsection{Scandium}

Scandium abundances begin above solar (about 0.15 dex) at the iron-poor end of the distributions
and then settle at [Fe/H]$\sim$$-$0.2\,dex. The distribution of scandium graphed in
Fig.\,\ref{fig2} basically mimics the trends of the $\alpha$-elements (in particular Si). 
The figures also show that there might be a slight upswing in scandium concentrations
for iron-rich stars, much like the upturn that potentially characterizes the
[Si/Fe] ratio. 

\citet[][]{Nis00} offered a thorough study of the abundances of scandium. The results
presented here for the metallicity range up to [Fe/H]$\sim$0.0 support their trends.
Their study stops at solar metallicities, and does not permit the confirmation of our interesting 
result for the more metal-rich stars. It is interesting to note, however, that in the [Sc/Fe] vs. [Fe/H]
plot presented by \citet[][]{Sad02} (combining their data with those of C00 and FG98) the same 
upward trend is suggested.

\subsection{Vanadium}

Despite the large scatter, the overall shape of the vanadium distribution resembles the functions for
silicon and scandium: overabundance at the iron-poor end, (almost) flat at solar values, and a potential
(but not clear) upturn for the most iron-rich stars. An overabundance of vanadium ($[$V$/$Fe$] \sim 0.2$ dex for iron-poor stars) was not detected by either FG98 or C00 (although the latter's plots hint at some trend), 
who proposed instead that the $[$V$/$Fe$]$ ratio followed iron for all metallicities. 
However, a look at Fig.\,19 of FG98 suggests a slight upper trend for higher 
metallicities as observed here.

It is important to note that vanadium seems to suffer strongly from NLTE effects (see Appendix).
For all metallicities, our analysis gives a decreasing [V/Fe] with increasing effective temperature trend.
This is probably the reason for the large scatter seen in Fig.\,\ref{fig2}. 
{As discussed for Ti, NLTE effects might e.g. be 
responsible (in part) for the decreasing trend seen for lower metallicity stars (Fe/H]$<$$-$0.2)}.

%, but is not
%responsible for the upturn witnessed at high metallicities, since in this domain of [Fe/H], we have
%stars within a large range of T$_{\mathrm{eff}}$. 

\subsection{Chromium}

Figs.\,\ref{fig2} and \ref{fig3} illustrate that the $[$Cr$/$Fe$]$ ratio stays fairly constant with increasing $[$Fe$/$H$]$, as was found by FG98 and C00.
For a given metallicity, chromium abundances have about half the star-to-star scatter encountered by C00. 
The authors noted that with a few weak
lines at their disposal, any dependence relation was tentative, and that the source of the scatter in
$[$Cr$/$Fe$]$ could not be readily ascribed to either systematic errors or cosmic effects. The C00
chromium distribution flirted with solar values whereas the stars in the current
survey have about 0.05 dex less chromium than the Sun. Although this shift is close
to the established error, it might conceal an underestimation if we suppose that the abundances of solar-type stars
should reflect solar values. Nevertheless, even if this indicates a systematic error, the trends
discussed should not be affected. As noted by FG98, the flatness is well explained by the balance between
the different types of sources for Fe and Cr \citep[][]{Tim95}.

\subsection{Manganese}

Manganese abundances increase with iron for the metallicity range studied, 
in agreement with \citet[][]{Nis00}, FG98, and \citet[][]{Sad02}. Figs.\,\ref{fig2} and \ref{fig3} show 
that $[$Mn$/$Fe$]$ begins with an underabundance of about $-0.2$ dex at [Fe/H]$\sim$$-$0.4, then rises 
to solar levels and beyond for the higher-metallicity stars in our sample. 

\subsection{Cobalt}

Hampered by a low number of measurable lines and possible strong NLTE effects (see Appendix), 
cobalt abundances possess
substantial star-to-star scatter, but the distribution appears to veer upwards for iron-rich stars, a result
similar to the one found by both FG98 and \citet[][]{Sad02} (a trend that recalls those of Si, Sc, and V
in our analysis). 
Given the dispersion in our data, we cannot
confirm the long plateau found by the former authors for [Fe/H] below solar. But our data insinuates 
decreasing [Co/Fe] as a function of [Fe/H] up to metallicities around solar {(which, as for Ti and
V, might in part be due to NLTE effects)}.

\subsection{Nickel}

Echoing the results of FG98 and C00, nickel abundances follow iron with low interstellar scatter per
metallicity (Fig. \ref{fig2}). The shape of the $[$Ni$/$Fe$]$ distribution exhibits a slight
decreasing trend (but probably a plateau) for $-0.6 < [$Fe$/$H$] < 0.0$ dex, and then an upturn. This upturn was also visible
in the data of FG98, and was discussed by C00. Furthermore, EAGLNT showed that their constant trend 
of $[$Ni$/$Fe$] = 0.0$ became disrupted and overabundance ensued once the stars passed the solar 
metallicity. 

It should be noted that the nickel abundances derived here are somewhat lower than
the values offered by the other authors. The reason for this might have to do with the determination of the
$\log{gf}$ values. However, as already noted above, the general trends should not perish from 
any systematic errors. A similar result might indeed be found in the study of FG98.

\section{Concluding Remarks}
\label{sec:conclusions}

In this paper, we have derived precise and uniform abundances for nine different elements (other than iron)
for a large sample of planet-host stars, as well as in a comparison sample of
stars without any discovered planetary-mass companions. 
The results were used to compare the two samples and to look for possible 
differences eventually connected to the presence of planets. The data was also 
used to explore galactic chemical evolution trends. 

Overall, we found that no significant differences were present between stars with and without
planetary companions, at least in the metallicity region where the two samples
overlap. The only elements showing potential trends were V, Mn, and to a 
lesser extent Ti and Co. {However, in no case were the differences clear.
These might be related e.g. to NLTE effects -- see Appendix\,\ref{appendix}}.

Furthermore, the available data gave us the possibility to investigate galactic chemical 
evolution trends for metal-rich stars with unprecedented detail. The results revealed some
interesting (and previously unnoticed) behavior concerning the metal-rich tail of the distributions 
(particularly for Si, Ca, and possibly Sc, and V).

The study of metal abundances in planet-host stars has already helped to 
clarify the formation processes of giant planets.
For example, they have shown that the efficiency of planetary formation is a strong function
of the metallicity \citep[e.g.][]{San01}. More details are likely to emerge as new
planets are found. The continuing study of the abundances in planet hosts might indeed
be a source of many more compelling results.
The determination of the abundances of volatile elements, for example, will give us the chance to
discuss the relative importance of differential accretion \citep[e.g.][]{Gon97,Smi01,Sad02}
in planet-host stars. Further interesting results might be derived from the study
of specific elemental abundance ratios like [C/O] \citep[e.g.][]{Gai00}.
On the other hand, some clues might come from the study of the 
sources of the elements present in planet-harboring stars. 
At present, we are working to extend the current study to other elements, as well as to
increase the number of stars in the current samples. 
Soon we hope to be able to answer 
many of the questions that were raised here.

\appendix{}
\section{Possible NLTE effects}
\label{appendix}

In this article, all the elemental abundances are derived assuming LTE. However, the dominance of this regime is
questionable in solar atmospheres. The assumption of LTE and a plane-parallel, homogeneous
model atmosphere can introduce systematic errors, and change the slope of elemental abundance
ratios $[$X$/$H$]$ versus $[$Fe$/$H$]$ \citep[e.g.][]{Edv93}. Plane-parallel atmospheres
can lead to too little adiabatic cooling and too much radiative heating \citep[e.g.][]{Che00}.

If, for metal-poor stars, the NLTE effects have been shown to be important \citep[e.g.][]{Edv93}, 
they are usually not very strong for their metal-rich counterparts \citep[][]{Edv93}, and 
NLTE corrections are frequently of the same order of magnitude (or lower) as the errors in the analysis.
\citet[][]{The99} have shown, for example, that the NLTE effects on iron abundances
for metal-rich dwarfs are very small. 

However, and at least for a few elements, the NLTE corrections seem to be quite strong.
For example, \citet[][]{Fel98} noticed that cool, metal-rich stars show up as underabundant in 
calcium, an effect attributed to NLTE effects, as previously discussed in \citet[][]{Dra91}. 
In this case, at least part of the errors were later shown to be due to wrongly-calculated damping 
parameters \citep[][]{Tho00}. But similar behavior was 
found by \citet[][]{Fel98} for nickel, and other elements studied by these authors also showed 
odd abundances. 

In a recent paper, \citet[][]{ThoFel00} showed that the use of a strict excitation equilibrium 
eliminates (at least in part) this problem. Following this result,
and since this is the method that we have used to estimate the effective temperatures for our 
program stars in \citet[][]{San00,San01,San02}, we should expect our parameters to be reasonably free from NLTE effects (i.e. NLTE corrections should not
be very strong).

\begin{figure}[t]
\psfig{width=\hsize,file=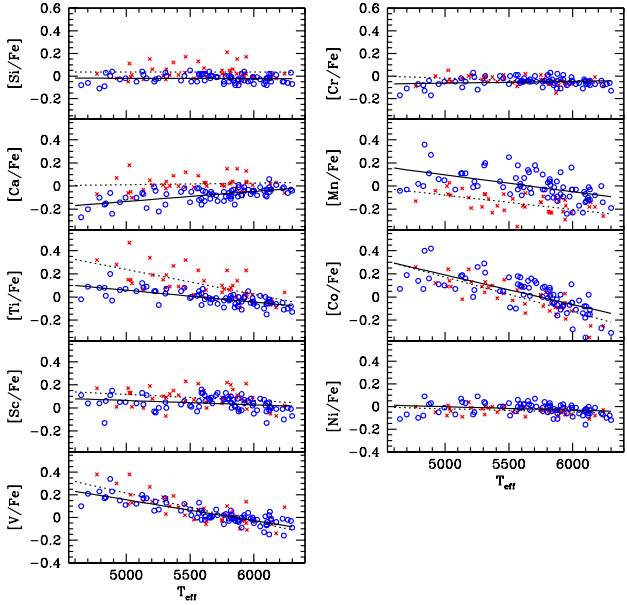}
\caption[]{[X/Fe] vs. T$_{\mathrm{eff}}$ plots for the nine elements studied in this paper.
The stars are separated based on their metallicities. The slopes represent linear 
least-squares fits to the iron-poor stars (crosses and dashed lines) and the iron-rich 
stars (circles and solid lines). }
\label{figa1}
\end{figure}

In Fig.\ref{figa1}, we plot the abundance ratios [X/Fe] for the nine elements studied in this paper
against T$_{\mathrm{eff}}$\footnote{We have done plots of the [Fe/H] abundances
as a function of T$_{\mathrm{eff}}$, $\log{g}$, and $\xi_t$, and have found no significant
trends. This gives us confidence about the reliability of our analysis, as expected since
for solar-metallicity stars neither \ion{Fe}{i} or \ion{Fe}{ii} (on which our parameter analysis relied \citep[][]{San00}) do not seem to suffer from
``important'' NLTE effects.}. In these plots, the stars are separated into two groups of metallicity:
objects with [Fe/H] lower than solar (circles) and above solar (disks). For each group,
a least-squares fit was done. The slopes are listed in Table\,\ref{tabNLTE}.

\begin{table}[ht] \centering
	\caption{ Slopes of $[$X$/$Fe$]$ ratios as functions of effective temperatures in dex per 1000 K. }
	\label{tabNLTE}
   	\begin{tabular}{lrr}	\hline
   	\noalign{\smallskip}
   	\noalign{\smallskip}
	Species	&	slope: $[$Fe$/$H$] < 0$	&	slope: $[$Fe$/$H$] \ge 0$	\\
	\noalign{\smallskip}
	\noalign{\smallskip}
	\hline
	\noalign{\smallskip}
	Si	&	$-0.001 \pm 0.028$	&	$-0.001 \pm 0.011$	\\
	Ca	&	$0.006 \pm 0.026$	&	$0.083 \pm 0.014$	\\
	Ti	&	$-0.213 \pm 0.041$	&	$-0.093 \pm 0.012$	\\
	Sc     	&	$-0.050 \pm 0.032$	&	$-0.031 \pm 0.016$	\\
	V	&	$-0.246 \pm 0.034$	&	$-0.176 \pm 0.014$	\\
	Cr	&	$-0.034 \pm 0.013$	&	$0.013 \pm 0.013$	\\
	Mn	&	$-0.107 \pm 0.030$	&	$-0.152 \pm 0.033$	\\
	Co	&	$-0.294 \pm 0.033$	&	$-0.243 \pm 0.033$	\\
	Ni	&	$-0.024 \pm 0.013$	&	$-0.025 \pm 0.015$      \\
	\noalign{\smallskip}
	\hline
   	\end{tabular}
\end{table}

As we can see from these plots, a few elements have considerable
dependence of the derived abundances on the effective temperature. 
If for Si, Sc, Cr, Ni, and Ca the effects seem to be very small, for
V, Ti and Co the difference between the K and F-dwarfs in our sample are of the order of 
0.2 to 0.3\,dex. The situation for Mn is intermediate.
Furthermore, for a few species we find relevant distinctions between the slopes for
iron-rich and iron-poor stars, namely for Ca and Ti. For the former of these
two elements, only the metal-rich dwarfs seem to be affected by a dependence on T$_{\mathrm{eff}}$,
whereas for Ti, the effect seems to be much stronger for the most metal-poor objects.
We note that the sensitivity of the \ion{Ti}{I} lines to NLTE effects has
already been discussed \citep[e.g.][]{Bro83}. A test with \ion{Ti}{ii} lines
has shown that the abundances obtained from these present a much smaller dispersion; the use of
\ion{Ti}{ii} might indeed be preferable (Shchukina 2002, personal communication).
In fact, since \ion{Ti}{ii} abundances are barely sensitive to T$_{\mathrm{eff}}$ \citep[][]{San00},
it is normal that no relation between [\ion{Ti}{ii}/Fe] and temperature is present.

The trends observed for V, Ti, Co, and Mn are reasons behind 
the large dispersions observed in the plots of Fig.\,\ref{fig2}. We note that for most
of the remaining elements the scatter is very small (except perhaps for Sc, for which
the cause may lie in the low number of spectral lines used).

Although we do not pretend to explain the causes for the observed trends here,
it is important to note them for future studies. In particular, for the elements
that seem to suffer more from these kind of effects, a precise
and reliable comparison between planetary hosts and non-host probably needs the use
of NLTE analysis.

\begin{acknowledgements}
  We wish to thank the Swiss National Science Foundation (Swiss NSF) 
  for the continuous support to this project. Support to N.C.S. from 
  Funda\c{c}\~ao para a Ci\^encia e Tecnologia, Portugal, 
  in the form of a scholarship is gratefully acknowledged. 
\end{acknowledgements}


\begin{thebibliography}{...}

\bibitem[Anders \& Grevesse(1989)]{And89} 
Anders E., \& Grevesse N., 1989, Geochim. et Cosmochim. Acta 53, 197

\bibitem[Brown et al.(1983)]{Bro83}
Brown J.A., Tomkin J., \& Lambert D.L., 1983, ApJ 265, L93

\bibitem[Chen et al.(2000)]{Che00} 
Chen Y.Q., Nissen P.E., Zhao G., Zhang H.W., \& Benoni T., 2000, A\&AS 141, 491 (C00)

\bibitem[Deliyannis et al.(2000)]{Del00}
Deliyannis C.P., Cunha K., King J.R., \& Boesgaard A.M., 2000,  AJ 119, 2437

\bibitem[Drake(1991)]{Dra91}
Drake J.J., 1991, MNRAS 251, 369

\bibitem[Edvardsson et al.(1993)]{Edv93}
Edvardsson B., Andersen J., Gustafsson B., et al., 1993, A\&A 275, 101 ())

\bibitem[Feltzing \& Gustafsson(1998)]{Fel98}
Feltzing S., \& Gustafsson B., 1998, A\&AS 129, 237 (FG98)

\bibitem[Fuhrmann et al.(1997)]{Fur97}
Fuhrmann K., Pfeiffer M.J., \& Bernkopf J., 1997, A\&A 326, 1081

\bibitem[Gaidos(2000)]{Gai00}
Gaidos E.J, 2000, Icarus 145, 637

\bibitem[Garc\'{\i}a L\'opez \& Perez de Taoro(1998)]{Gar98}
Garc\'{\i}a L\'opez R., \& Perez de Taoro M.R., 1998, A\&A 334, 599

\bibitem[Gonzalez et al.(2001)]{Gon01} 
Gonzalez G., Laws C., Tyagi S., \& Reddy B.E., 2001, AJ 121, 432

\bibitem[Gonzalez \& Laws(2000)]{Gon00} 
Gonzalez G., \& Laws C., 2000, AJ 119, 390

\bibitem[Gonzalez(1998)]{Gon98}
Gonzalez, G. 1998, A\&A, 334, 221

\bibitem[Gonzalez(1997)]{Gon97}
Gonzalez G., 1997, MNRAS 285, 403

\bibitem[Israelian et al.(2003)]{Isr03}
Israelian G., Santos N.C., Mayor M., \& Rebolo R., 2003, A\&A, submitted
 
\bibitem[Israelian et al.(2001)]{Isr01}
Israelian G., Santos N.C., Mayor M., \& Rebolo R., 2001, Nature 411, 163

\bibitem[Jones et al.(2002)]{Jon02}
Jones H., Butler P., Tinney C., et al., 2002, MNRAS 333, 871

\bibitem[Kurucz(1993)]{Kur93} 
Kurucz R. L., 1993, CD-ROMs, ATLAS9 Stellar
Atmospheres Programs and 2~${\rm km}~{\rm s}^{-1}$ Grid
(Cambridge: Smithsonian Astrophys. Obs.)

\bibitem[Kurucz et al.(1984)]{Kur84} 
Kurucz R.L., Furenlid I, Brault J., \& Testerman L., 1984, Solar Flux Atlas from 296 to 1300 nm, NOAO Atlas No. 1 

\bibitem[Nissen et al.(2000)]{Nis00}
Nissen P.E., Chen Y.Q., Schuster W.J., \& Zhao G., 2000, A\&A 353, 722

\bibitem[Reddy et al.(2002)]{Red02}
Reddy B.E., Lambert D.L., Laws C., Gonzalez G., \& Covey K., 2002, MNRAS 335, 1005

\bibitem[Reid(2002)]{Rei02}
Reid I.N., 2002, PASP 114, 306

\bibitem[Ryan(2000)]{Rya00}
Ryan S.G., 2000, MNRAS 316, L35

\bibitem[Sadakane et al.(2002)]{Sad02}
Sadakane K., Ohkubo M., Takada Y., et al., 2002, PASJ, in press

\bibitem[Sadakane et al.(2001)]{Sad01}
Sadakane K., Ohkubo M., Sato S., et al., 2001, PASJ 53, 315

\bibitem[Sadakane et al.(1999)]{Sad99}
Sadakane K, Honda S., Kawanomoto S., Takeda Y, \& Takada-Hidai M., 1999, PASJ 51, 505

\bibitem[Santos et al.(2003)]{San02} 
Santos N.C., Israelian G, Mayor M., Rebolo R., \& Udry S., 2003, A\&A 398, 363

\bibitem[Santos et al.(2002)]{San02b}
Santos N.C., Garcia Lopez R.J., Israelian G., et al., 2002, A\&A 386, 1028

\bibitem[Santos et al.(2001)]{San01} 
Santos N.C., Israelian G, \& Mayor M. 2001, A\&A, 373, 1019 

\bibitem[Santos et al.(2000)]{San00} 
Santos N.C., Israelian G., \& Mayor M. 2000, A\&A 363, 228 

\bibitem[Smith et al.(2001)]{Smi01} 
Smith V.V., Cunha C., \& Lazzaro D., 2001, AJ 121, 3207

\bibitem[Sneden(1973)]{Sne73} 
Sneden C., 1973, Ph.D. thesis, University of Texas

\bibitem[Th\'evenin \& Idiart(1999)]{The99}
Th\'evenin F., \& Idiart T.P., 1999, ApJ 521, 753

\bibitem[Thoren(2000)]{Tho00}
Thoren P., 2000, A\&A 358, L21

\bibitem[Thoren \& Feltzing(2000)]{ThoFel00}
Thoren P., \& Feltzing S., 2000, A\&A 363, 692

\bibitem[Timmes et al.(1995)]{Tim95}
Timmes F.X., Woosley S.E., \& Weaver T.A., 1995, ApJS 98, 617

\bibitem[Wasson(1985)]{Was85}
Wasson J.T., 1985. In: ``Meteorites: Their Record of Early Solar System History", W.H. Freeman 
Publishers, USA

\bibitem[Zhao et al.(2002)]{Zha02}
Zhao G., Chen Y.Q., Qiu H.M., \& Li Z.W., 2002, AJ 124, 2224


\end{thebibliography}
\end{document}